\documentclass[useAMS,usenatbib,usegraphicx]{mn2e}
\usepackage{pslatex}
\usepackage{amsmath,amsfonts}
\newcommand{\Hfull}{\mathcal{H}}
\newcommand{\Hsec}{\mathcal{H}_{\idm{sec}}}

\newcommand{\Hpert}{\mathcal{H}_{\idm{pert}}}
\newcommand{\HpertJ}{\mathcal{H}^J_{\idm{pert}}}
\newcommand{\HKepl}{\mathcal{H}_{\idm{kepl}}}
\newcommand{\Hint}{\mathcal{H}_{\idm{int}}}

\newcommand{\R}{\mathcal{R}}
\def\vec#1{{\mathbf{#1}}}
\def\idm#1{{\mbox{\scriptsize #1}}}
\def\rev#1{{#1}}
\title[Secular planetary problem]
{A secular theory of coplanar, non-resonant planetary system}
\author[C. Migaszewski and K. Go\'zdziewski]{Cezary Migaszewski$^{1}$\thanks{E-mail:
c.migaszewski@astri.uni.torun.pl} and Krzysztof Go\'zdziewski$^{1}$\footnotemark[1]\thanks{E-mail:
k.gozdziewski@astri.uni.torun.pl}\\
$^{1}$Toru\'n Centre for Astronomy, Nicolaus Copernicus University, 
Gagarin Str. 11, 87-100 Toru\'n, Poland}
\begin{document}
\date{Accepted 2008 May 7.  Received 2008 May 2; in original form 2008 February 1}
\pagerange{\pageref{firstpage}--\pageref{lastpage}} \pubyear{2008}
\maketitle
\label{firstpage}
\begin{abstract}
We present the secular theory of coplanar $N$-planet system, in the absence of
mean motion resonances between the planets. This theory relies on the averaging 
of a perturbation to the two-body problem over the mean longitudes.  We expand
the perturbing Hamiltonian in Taylor series with respect to the ratios of
semi-major axes  which are considered as small parameters, without  direct
restrictions on the eccentricities. Next, we average out the resulting series
term by term. This is possible thanks to a particular but in fact quite
elementary choice of the integration variables. It makes it possible to avoid
Fourier expansions of the perturbing Hamiltonian. We derive high order
expansions of the averaged secular Hamiltonian (here, up to the order of 24)
with respect to the semi-major axes ratio. The resulting secular theory is a
generalization of the octupole theory.  The analytical results are compared with
the results of numerical (i.e., practically exact) averaging. We estimate
the convergence radius of the derived expansions, and we propose a further
improvement of the algorithm. As a particular application of the method, we 
consider the  secular dynamics of three-planet coplanar system. We focus on
stationary solutions in the HD~37124 planetary system.
\end{abstract}


\section{Introduction}
The recent discoveries of extrasolar planetary systems bring new and interesting
problems regarding their dynamical stability and long-term evolution. At
present, at least 30 multi-planet systems are known and their number is still
growing, thanks to refined techniques of observations. Surprisingly, the orbital
parameters of these systems are very different from those typical in the Solar
System  architecture --- large planetary masses and  eccentricities are common.
Simultaneously, these systems usually are compact. Likely, this property is a
consequence of  the observational selection. The most effective detection
techniques, like the radial velocity observations, rely on indirect effects of
mutual interactions between  planets and their host star. Many of the known
multi-planet systems are supposed to be involved in short-term mean motion
resonances (MMRs). However, there are also configurations with relatively well
separated orbits. In that case the secular interactions may lead to interesting
dynamical phenomena.

\rev{
To study the long term dynamics of planetary systems, different analytical and
numerical techniques are used. The analytical approach is much more effective in
the investigations of global, qualitative dynamics than widely applied numerical
techniques (including fast indicators or direct numerical integration of the
equations of motion). The numerical experiments provide only limited (or local)
information on the dynamical features of the studied configurations. Usually,
the interpretation of the results of massive calculations can be problematic
without solid theoretical background. In contrast, analytical techniques offer
much deeper insight into qualitative properties of motion. The analytical
approach makes it possible to explore large volume of the phase space. This is
crucial for the dynamical studies of extrasolar planetary systems detected
during short time of observations. These observations have relatively large
errors, they are typically irregularly sampled or degenerated (in the sense that
they can provide only limited information on the system state, like the radial
velocity technique). This leads to poorly determined or unconstrained orbital
and physical parameters of the detected systems. In that case, the analytical
theories help us to investigate and/or to detect global properties of the
solutions to the equations of motion in  observationally permitted ranges of the
parameters. We can investigate in detail certain families of these solutions,
their bifurcations and stability. Examples of such solutions are stationary
solutions (equilibria) or periodic orbits that build a skeleton of the phase
space. Investigating these families, we follow the classic methodology invented by
Poincar\'e.  The analytical theories are milestones for detailed
numerical studies of particular aspects of the dynamics. Hence, their constant
development is always desirable.
}

Moreover, due to extreme parameters of the studied configurations, the classic
planetary theory developed so far is often too week. For instance, the classic
Lagrange-Laplace theory \citep{Murray2000} designed as a model of the secular
dynamics of planets in the Solar System fails in the case of large
eccentricities and inclinations. Hence, new analytical  and semi-analytical
theories are recently developed, breaking the limitations of the classic
approach.  One of the most effective techniques for studying the secular 
dynamics of extrasolar systems  has been recently invented by
\cite{Michtchenko2004} and further developed in \citep{Michtchenko2006}. These
papers are devoted to a study of two-planet configurations. In this work, we
consider an analytical secular theory of a coplanar system of $N$-planets (point
masses) under assumption that the  orbital configurations are not involved in
strong mean-motion resonances and that they are far from collisions zones of
orbits. We calculated the averaged perturbation in the form of  power series
with respect to the semi-major axes ratios up to very high order (equal to 24 in
the present work). These expansions have no  explicit limits on the
eccentricities provided that the non-resonance condition is satisfied. Our
development is elementary and is based on very basic properties of the Keplerian
motion.  Although  it concerns the two-planet system, we show that it can be
easily generalized for the case of $N$-planet configurations. Hence, the theory
can be regarded  not only as an attempt to improve the secular theories for
two-planet systems in
\citep[e.g.,][]{Gallardo2005,Henrard2005,Libert2005,Libert2006,Ji2007,Veras2007},
relying on the classic expansion of the perturbing Hamiltonian in eccentricities
\citep{Murray2000,Ellis2000} or the octupole theory that makes use on the
averaging of the low-order expansion of the perturbing function in the
semi-major axes ratio \citep{Ford2000,Blaes2002,Lee2003}.  We try to reduce the
limitations of the classic theory of non-resonant systems. 

The plan of this paper is as follows. In Section~2, we describe a general model
of a coplanar configuration of $N$ planets. We introduce  the expansion of
perturbing Hamiltonian and a very simple and basic algorithm of its averaging.
We compare the results of the method with the  outcome of the octupole theory,
and we discuss some subtle differences between these theories. We also present
the results of the tests of the expansion, taking as examples a few known
multi-planet configurations which apparently fit well in the framework of the
non-resonant secular theory. The exact semi-numerical method is helpful to
determine absolute bounds of the validity of the analytic approach. We also
outline a further improvement of the averaging algorithm.
In Section~3 we
construct the secular model of the three-planet system and we perform a
preliminary study of the secular dynamics of a few known three-planet
configurations. In particular, we focus on the equilibria in the secular
problem,
and we found interesting stationary solutions in the extrasolar system
HD~37124~\citep{Vogt2005,Gozdziewski2007}. 

\section{The secular dynamics of a multi-planet system}
The Hamiltonian of a multi-planet system with respect to canonical Poincar\'e 
variables \citep[see, e.g.,][]{Laskar1995,Michtchenko2004}  can be expressed by
a sum of two terms,
\begin{equation}
\Hfull = \HKepl + \Hpert,
\label{Hfull}
\end{equation}
where
\begin{equation}
\HKepl = \sum_{i=1}^{N} {\bigg( \frac{\mathbf{p}_i^2}{2 \beta_i} 
- \frac{\mu_i \beta_i}{r_i} \bigg)}
\label{HKepl}
\end{equation}
is for the integrable part comprising of the direct sum of the relative,
Keplerian motions of $N$ planets and the host star. Here, the dominant point
mass of the star is ${m_0}$, and $m_i \ll m_0$, $i=1,\ldots,N$ are the point
masses of the $N$-planets.  For each planet--star pair we define the mass
parameter ${\mu_i=k^2~(m_0+m_i)}$ where $k$ is the Gauss gravitational constant,
and ${\beta_i=(1/m_i+1/m_0)^{-1}}$ are the so called reduced masses. Due to
mutual interactions between the planets, the Keplerian part is perturbed by a
function $\Hpert$, 
\begin{equation}
\Hpert \equiv \R = \sum_{i=1}^{N-1} \sum_{j>i}^{N} {\bigg(
 - \underbrace{\frac{k^2 m_i m_j}{\Delta_{i,j}}}_{\textrm{\small direct part}} +
\underbrace{\frac{\mathbf{p}_i \cdot \mathbf{p}_j}{m_0}}_{\textrm{\small indirect
part}}\bigg)},
\label{Hpert}
\end{equation}
where  ${\mathbf{r}_i}$ are for the position vectors of the planets relative to
the star,  ${\mathbf{p}_i}$ are for their conjugate momenta relative to the {\em
barycenter} of the whole $(N+1)$-body system, 
${\Delta_{i,j}=\|\mathbf{r}_i-\mathbf{r}_j\|}$ denote the relative distance
between planets $i$  and $j$. It is well known that even in the simplest case of
three  point masses (the star and two mutually  interacting planets), the
problem is non-integrable and is not possible to obtain its exact analytical
solutions. In practice, such solutions can be only derived in the form of
approximations  derived by means of different perturbation techniques
\citep[e.g.,][]{Murray2000,Morbidelli2003,FerrazMello2007}. 

To apply the canonical perturbation theory, we first transform $\Hfull$ to the
following form:
\begin{equation}
 \Hfull(\vec{I},\vec{\phi}) = \HKepl(\vec{I}) + 
 \Hpert(\vec{I},\vec{\phi}),
\label{eq:poincare} 
\end{equation}
where $(\vec{I},\vec{\phi})$ stand for the action-angle variables, and
$\Hpert(\vec{I},\vec{\phi}) \sim \epsilon \HKepl(\vec{I})$,  where $\epsilon \ll
1$ is a small parameter. In the absence of this perturbation, the system is 
trivially integrable. However, with the perturbation added, the dynamics of the 
full system become extremely complex.  In the realm of the Hamiltonian canonical
theory, the approximate, analytical solutions to this problem may be derived by
an expansion of the perturbation with respect to the small parameter and by
subsequent simplification of the lowest order terms by means of appropriate
canonical contact transformations. This idea of Delaunay  appears in many
``incarnations''. One of its first novel realizations is known as the von~Zeipel
method \citep[e.g.,][]{Brumberg1995}.  Much more improved version of this
technique that not require series inversion  has been invented by
\cite{Hori1966} and next refined by \cite{Deprit1969}.  This theory is well
known in the literature as the Lie-Hori-Deprit method. For an excellent review
of these methods see a monograph by \cite{FerrazMello2007}. Another method of
seeking for approximate solutions to Eq.~\ref{eq:poincare} relies directly on
the averaging proposition \citep[see, e.g.,][]{Arnold1993}. Usually, the
canonical angles $\vec{\phi}$ can be divided onto two classes: {\em fast} and
{\em slow} ones. By averaging the perturbing part with respect to the fast
angles over their periods, we obtain the secular perturbation Hamiltonian which
does not depend on these fast angles. Simultaneously,  their conjugate momenta
become integrals of the secular problem. In the planetary model with a dominant
stellar mass, we have two natural time-scales of motion: the orbital motion of
the planets and a slow evolution of their orbits. Assuming that no strong  mean
motion resonances are present, and the system is far enough from collisions, 
the averaging makes it possible to reduce the number of the degrees of freedom
and to obtain qualitative information on the long-term changes of the slowly
varying orbital elements (i.e., on the slow angles and their conjugate momenta).

To apply each one of these methods,  the Hamiltonian of the $N$-planet system
should be first transformed to the required form, Eq.~\ref{eq:poincare}. This
can be accomplished by expressing it with  respect to the canonical Poincar\'e
elements \citep{Murray2000,Michtchenko2004}:
\begin{eqnarray}
{l_i \equiv \lambda_i}, & \quad {L_i=\beta_i~\sqrt{\mu_i~a_i}},\nonumber\\
{g_i \equiv -\varpi_i}, & \quad  {G_i=L_i~(1 - \sqrt{1-e_i^2})},\\
{h_i \equiv -\Omega_i}, & \quad{H_i=L_i \sqrt{1-e_i^2}~(1 - \cos~I_i)},\nonumber
\label{delaunay}
\end{eqnarray}
where $\lambda_i$ are the mean longitudes, $a_i$ stand for canonical semi-major
axes,  $e_i$ are for the eccentricities,  $I_i$ denote inclinations, 
\rev{$\varpi_i$ are the longitudes of pericenter},  and $\Omega_i$ denote the
longitudes of the  ascending node. We note that 
\rev{
the transformation between the canonical orbital elements of Poincar\'e, $a_i$,
$e_i$, $I_i$, $\varpi_i$, $\Omega_i$  and associated  Cartesian coordinates and
momenta may be derived by the formal two-body transformation between classic
(astro-centric) Keplerian elements and the Cartesian coordinates
\citep[e.g.,][]{Morbidelli2003,Ferrazmello2006}.
}
Moreover, in the settings adopted here, the rectangular coordinates and momenta
are understood  through the Cartesian positions of planets relative to the star,
and, according to the definition of the Poincar\'e variables, the canonical
momenta are taken relative to the {\em barycenter} of the system. 

The $N$-body Hamiltonian expressed in terms of the Poincar\'e variables has the
form of:
\[
\Hfull = -\sum_{i=1}^N \frac{\mu_i^2 \beta^3_i}{2 L_i^2} +
\Hpert\underbrace{(L_i,l_i,G_i,g_i,H_i,h_i)}_{i=1,\ldots,N}.
\]
In this Hamiltonian, $l_i$ play the role of the fast angles. In the absence of
strong MMRs,  these angles can be eliminated by the following
averaging formulae:
\begin{equation}
\Hsec=\frac{1}{(2 \pi)^N}\underbrace{\int_0^{2 \pi} \ldots 
\int_0^{2\pi}}_{i=1,\ldots,N}{\Hpert \, d\lambda_1 \ldots d\lambda_N}.
\label{eq:secular}
\end{equation}
(As we see below, it can be also applied for some selected pairs of planets).
Hence, the conjugate momenta $L_i$ become integrals of the secular model and the
Keplerian part is a constant that does not contribute to the equations of
motion. However, the calculation of the multiple integral is quite a difficult
task which is a central part of the problem.  We try to solve it with quite
basic mathematical properties of the Keplerian osculating orbits.

\rev{
We should keep in mind that the averaging of the secular Hamiltonian in the
problem over {\em Keplerian} motions implies truncation of the perturbation
to first order in the masses (more generally, to the first order in the
perturbation parameter $\epsilon$). For large mutually interacting planets
(or binary stars), the deviations of the true orbits from the Keplerian
approximation during the orbital period may become significant, and the
secular theory may fail. Nevertheless, it is a common drawback of the idea
behind Eq.~{\ref{eq:secular}}. For the same reason, the classic perturbation
techniques usually fail in the case of close encounters between the planets. 
In such an instance, some other criteria helping to explore the stability
regions may be applied, for example, the Hill stability criterion
\citep{Marchal1982,Gladman1993,Barnes2006,Michtchenko2008,Barnes2008}.
}

\subsection{The indirect part of the disturbing function}
We start with the averaging of the {\em indirect} part of the disturbing
Hamiltonian.
\rev{
The result of this averaging  can be found in \cite{Brouwer1961}, nevertheless,
to make this paper self-consistent, we present
the calculations in detail. 
}
The indirect part
is a scalar product of canonical momenta $\mathbf{p}_i$, which
have the form of:
\begin{equation}
\mathbf{p}_i = {\beta'}_i \mathbf{\dot{r}}_i - \sum_{j \neq i} {\frac{m_i m_j}{M}
\mathbf{\dot{r}}_j},
\end{equation}
where ${\beta'}_i = \left[1/m_i + 1/(M-m_i)\right]^{-1}$ and $M$ is the  total
mass of the system. The scalar product $\mathbf{p}_i \cdot \mathbf{p}_j$ 
includes terms of the type of $\mathbf{\dot{r}}_i \cdot \mathbf{\dot{r}}_j$.
Moreover,  each product $\mathbf{p}_i \cdot \mathbf{p}_j$ depends on all 
astro-centric velocities of the planets,  $\mathbf{\dot{r}}_i$ ($i=1,\ldots,N$).
Apparently, to average out the indirect part of the disturbing function, we must
compute multiple integral over all mean longitudes $\lambda_i$ ($i=1,\ldots,N$).
In fact, this integral can be reduced to a sum of double integrals computed for
all pairs of planets, i.e., we average out expressions of the form of 
$\mathbf{\dot{r}}_i \cdot \mathbf{\dot{r}}_j$. The result is the following:
\begin{equation}
\frac{1}{(2\pi)^2} \int_{0}^{2 \pi} \int_{0}^{2 \pi} {\mathbf{\dot{r}}_i
\cdot \mathbf{\dot{r}}_j ~d\mathcal{M}_i d\mathcal{M}_j} = 
\delta_{i,j}~a_i^2~n_i^2,
\end{equation}
where $n_i$ denote mean motions of the planets and $\delta_{i,j}$   stands for
the Kronecker delta.   Note, that the averaging over the mean anomalies gives
the same results as the averaging over  the mean  longitudes under the condition
that the integration limits are set to  $0$ and $2\pi$, respectively. Clearly,
the indirect part of the disturbing function does not contribute to  the secular
dynamics of the system  because it depends on $L_i$ only \citep{Brouwer1961},
\cite[see also][]{Michtchenko2004}. We note that this result is exact as far as
the assumptions of the averaging principle are fulfilled (we are far enough from
the MMRs and there are present  two different time-scales in
the problem).

\subsection{The direct part of the disturbing function}
Now we have a more difficult problem to resolve. For the $N$-planet system, the
averaged  direct part of the disturbing function has the form of:
\begin{equation}
\Hsec =
\sum_{i=1}^{N-1}  \sum_{j>i}^{N}{\Hsec^{(i,j)}},
\label{eq:secular2}
\end{equation}
where the multiple integral Eq.~(\ref{eq:secular})  over all mean anomalies is
reduced to a sum of secular Hamiltonians describing mutual interactions between
all pairs of planets; i.e., for each pair $(i,j)$,  where $i<j$ and $a_i<a_j$,
we have:
\begin{equation}
\Hsec^{(i,j)} = 
 \frac{1}{(2\pi)^2} \int_0^{2\pi} \int_0^{2\pi} 
-{\frac{k^2 m_i m_j}{\Delta_{i,j}} d\mathcal{M}_i d\mathcal{M}_j}.
\label{eq:rsec}
\end{equation}
Hence, the secular model of $N$-planet system can be reduced to a simple sum of
two-planet Hamiltonians.  

Now, we compute the double integral for a fixed pair of planets  $i$ and $j$.
The distance between these planets is the following:
\begin{equation}
\Delta_{i,j} = \sqrt{r_i^2 + r_j^2 - 2 r_i r_j \cos{\psi_{i,j}}},
\label{squareroot1}
\end{equation}
where $\psi_{i,j}$ is the angle between vectors $\mathbf{r}_i$ and
$\mathbf{r}_j$:
\begin{equation}
\cos{\psi_{i,j}} = \frac{\mathbf{r}_i \cdot \mathbf{r}_j}{r_i r_j} = 
\frac{x_i x_j + y_i y_j}{r_i r_j}.
\end{equation}
This formulae can be rewritten to:
\begin{equation}
\Delta_{i,j} = r_j \sqrt{1 - 2 \frac{1}{r_j} \Big(x_i \frac{x_j}{r_j} 
+ y_i\frac{y_j}{r_j}\Big) + 
\Big(\frac{r_i}{r_j}\Big)^2}.
\label{eq:squareroot2}
\end{equation}
According to the Kepler problem theory, $r_i$ and $r_j$ may be expressed through
the eccentric anomaly, $E$, or, equivalently, through the true anomaly, $f$. We
write down appropriate expressions  for planet $i$ and $j$, respectively: 
\[
r_i = a_i (1 - e_i \cos{E_i}), \quad
r_j = \frac{a_j (1 - e_j^2)}{1 + e_j \cos{f_j}}.
\]
Here, $E_i$ is the eccentric anomaly of the inner planet in the
selected pair of interacting bodies,  and $f_j$ is the true anomaly 
of the outer planet. 
Moreover:
\[
\frac{x_j}{r_j} = \cos\left({f_j + \Delta{\varpi_{i,j}}}\right), 
\quad \frac{y_j}{r_j} = \sin\left({f_j + \Delta{\varpi_{i,j}}}\right),
\]
where $\Delta\varpi_{i,j} \equiv \left(\varpi_j - \varpi_i\right)$,
and
\[
x_i = a_i (\cos{E}_i - e_i), \quad y_i = a_i \sqrt{1 - e_i^2} \sin{E_i}.
\]
The dependence of the {\em two-body formulae} on $\Delta\varpi_{i,j}$ may seem
strange. However, when we investigate the co-planar system of two  particular
planets, we are free to choose the reference frame
because the mutual interaction between these planets depend only on their  {\em
relative} orbital phases.  To be more specific, we must calculate the distance
between planets $i$ and $j$, or
the scalar product $\vec{r}_i \cdot \vec{r}_j \equiv r_i r_j
\cos{\psi_{i,j}}$ (Eq. \ref{squareroot1}). That is obviously independent on the
reference frame. In general, we could write: 
\begin{equation} 
\vec{r}_i \cdot \vec{r}_j  =
{\mathbb A}_i \vec{r}_i\big|_{{\cal F}_i} 
\cdot {\mathbb A}_j \vec{r}_j\big|_{{\cal F}_j}, 
\end{equation} 
where ${\cal F}_{i,j}$ are the orbital reference frames  of the inner and outer
planet, respectively,  matrices ${\mathbb A}_i, {\mathbb A}_j$ represent
Eulerian rotations of ${\cal F}_{i,j}$ to a common reference frame (${\cal
F}$) for both orbits. Because this  frame may be chosen freely,  we fix the
$x$-direction of the common frame along the apsidal line of the inner planet
(still, for a particular pair of planets).  Then ${\mathbb A}_i \equiv
{\mathbb E}$ and ${{\cal F}_j}$ must be rotated by angle $\Delta{\varpi_{i,j}}$.
That can be repeated for each pair of planets in multi-planet system because the
secular Hamiltonian is represented by a sum of formally independent two-planet
terms.

Finally, the inverted distance between planets $i,j$ can be expressed
as follows:
\[
\frac{1}{\Delta_{i,j}} = \frac{1 + e_j \cos{f_j}}{a_j (1 - e_j^2)} 
\Big[A \alpha_{i,j}^2 -2 B \alpha_{i,j} +1\Big]^{-1/2},
\]
where $ \alpha_{i,j} \equiv {a_i}/{a_j} < 1$, and
\begin{eqnarray}
A &\equiv& \frac{(1 + e_j \cos{f_j})^2 (1 - e_i \cos{E_i})^2}{(1 - e_j^2)^2}, \\
B &\equiv&  \frac{1 + e_j \cos{f_j}}{1 - e_j^2} \Big[(\cos{E_i} - e_i) 
\cos\left({f_j + \Delta{\varpi_{i,j}}}\right) +\\
&+& \sqrt{1-e_i^2} \sin{E_i} \sin\left({f_j + \Delta{\varpi_{i,j}}}\right)
\Big] \nonumber.
\end{eqnarray}
Now we underline that the position of the inner planet is  given through the
{\em eccentric} anomaly while the position of the outer planet  is given with
respect to the {\em true} anomaly. The formulae under the square root are
expressed by a polynomial of trigonometric functions and, as we can see below,
that is critically important property making it possible to calculate  the
integral in Eq.~\ref{eq:rsec}.

Now, we expand the inverse of the distance between planets $i$ and $j$ in  
Taylor series with respect to small parameter $\alpha_{i,j}$. The series are
evaluated around $\alpha_{i,j}=0$ as follows:
\begin{equation}
\frac{1}{\Delta_{i,j}} = \frac{1 + e_j \cos{f_j}}{a_j (1 - e_j^2)} \sum_{l=0}^{\infty}
\bigg[\frac{1}{l!} \frac{d^l \mathcal{D}}{d\alpha_{i,j}^l} \bigg|_{\alpha_{i,j}=0} 
\alpha_{i,j}^l \bigg],
\label{eq:oneoverD}
\end{equation}
where
\begin{equation}
\mathcal{D} = \Big[A \alpha_{i,j}^2 -2 B \alpha_{i,j} +1\Big]^{-1/2}.
\label{eq:eqf}
\end{equation}
As the final result of this  expansion, we obtain a polynomial with respect to 
trigonometric functions of the anomalies which has the general form of:
\begin{equation}
\frac{1}{\Delta_{i,j}} = \sum_{\mathbf{p}} \Big[C_{\mathbf{p}} (\cos{E_i})^{p_1} 
(\sin{E_i})^{p_2} 
(\cos{f_j})^{p_3} (\sin{f_j})^{p_4}\Big].
\label{eq:oneoverDgeneral}
\end{equation}
Here, $\vec{p} \equiv (p_1,p_2,p_3,p_4) \in {\mathbb N}^4$  is a vector of
natural numbers and  $C_{\vec{p}}$  are coefficients depending on eccentricities
and semi-major axes. One more step is still necessary. The integral in 
Eq.~\ref{eq:rsec} must be computed with respect to the {\em mean} anomalies. 
Here, the classic theories make use on $\sin f$ and $\cos f$ expressed through
Fourier series of  the mean anomalies with coefficients dependent on the
eccentricities. 

However, we found that it is possible to avoid these expansions. Again, using the basic formulae
of the Keplerian motion, we perform a formal change of variables in Eq.~\ref{eq:rsec} with:
$d\,\mathcal{M}_i = \mathcal{I}_i dE_i$ and $d\,\mathcal{M}_j = \mathcal{J}_j df_j$, where
functions $\mathcal{I}_i \equiv  \mathcal{I}_i(E_i,e_i)$  and $\mathcal{J}_j \equiv
\mathcal{J}_j(f_j,e_j)$ are defined with:
\begin{equation}
\mathcal{I}_i(E_i,e_i) = 1 - e_i \cos{E_i}, \quad
\mathcal{J}_j(f_j,e_j) = \frac{\left(1-e_j^2\right)^{3/2}}{\left(1 + e_j \cos{f_j}\right)^2}.
\label{eq:ij}
\end{equation}
After this change of variables, the average in Eq.~\ref{eq:rsec} is equivalent to
calculation of the following double integral:
\begin{equation}
\Hsec^{(i,j)} = 
 \frac{1}{(2\pi)^2} \int_0^{2\pi} \int_0^{2\pi} 
                  -k^2 m_i m_j \frac{1}{\Delta_{i,j}} \mathcal{I}_i(E_i,e_i) 
   \mathcal{J}_j(f_j,e_j) \, d E_i \, d f_j,
\label{eq:rsecl}
\end{equation}
where 
${\Delta^{-1}_{i,j}} \equiv \Delta^{-1}_{i,j}(a_i,a_j,e_i,e_j,E_i,f_j)$.

Fortunately, functions $\mathcal{I}_i$ are again  trigonometric polynomials  and they do
not change the general, polynomial form  of Eq.~\ref{eq:oneoverDgeneral}.
However, the second scaling function is not a polynomial with respect to
$\cos{f_i}$ or $\sin{f_i}$.  Yet Eq.~\ref{eq:oneoverD} involves a factor 
$\left(1 + e_j \cos{f_j}\right)$. It cancels out one power of $\left(1 + e_j
\cos{f_j}\right)$ appearing in the denominator of Eq.~\ref{eq:ij}. In order to
calculate the expansion in Eq.~\ref{eq:oneoverD}, for $l>0$ we differentiate
${\cal D}(\alpha_{i,j})$  with respect to $\alpha_{i,j}$ as the composite
function (Eq.~\ref{eq:eqf}). This operation emerges factors  of the type of $A^r
B^s$, where $n = r + s \geq 1$. Looking at the  general form of $A$ and $B$ we
see that the term $\left(1 + e_j \cos{f_j}\right)$  appears with  natural powers
larger than $1$ and it cancels out  remaining $\left(1 + e_j \cos{f_j}\right)$
in the denominator of Eq.~\ref{eq:ij}. In this way, the general form of
trigonometric polynomial in Eq.~\ref{eq:oneoverDgeneral} is preserved. Still,
the free term in the Taylor expansion of ${\cal D}$ leads to an expression
involving $\left(1 + e_j \cos{f_j}\right)^{-1}$. Fortunately, we must integrate
such term with limits from $0$ to $2\pi$ and this effectively can be reduced to
averaging out $r_j$ over the orbital period (or the 
whole range of the true anomaly).

Finally, we can integrate Eq.~\ref{eq:rsecl} term by term.  Basically, the
problem has been reduced to  the calculation of definite integrals from products
of trigonometric functions $\sin(x)$ and $\cos(x)$ in some natural powers. These
integrals can be derived quite easily, at least in principle, nevertheless with
increasing order of the Taylor expansion, the calculations become extremely
tedious. To accomplish them, we used MATHEMATICA and fast AMD-Opteron computer.

The final result of the averaging is an expansion of the secular, two-body
Hamiltonian for a chosen pair of planets $i$ and $j$:
\begin{eqnarray}
& & \Hsec^{(i,j)} = -\frac{k^2 m_i m_j}{a_j} \times  \nonumber \\
&& \quad \times \left[1 + \sqrt{1-e_j^2} \sum_{l=2}^{\infty}
{\left(\frac{\alpha_{i,j}}{1-e_j^2}\right)^l
\mathcal{R}^{(i,j)}_l(e_i,e_j,\Delta\varpi_{i,j})}\right].
\label{expansion}
\end{eqnarray}
Looking at the general form of this secular Hamiltonian, we learn that the role
of a formal parameter in the power-series  expansion of $\Hsec$ plays
(apparently) the following expression: 
\begin{equation} X_{i,j} \equiv \frac{\alpha_{i,j}}{1-e_j^2} =
\frac{a_i}{a_j (1-e_j^2)}. 
\end{equation} 
Obviously, these series cannot
converge if $X_{i,j} \geq 1$, hence we require that $X_{i,j}<1$.  Of course,
this is only the necessary (and as we see below, very rough) condition for the
convergence of these series. 
\rev{
However, as we explain in Sect.~2.5, attributing to $X_{i,j}$ 
the role of a 
parameter deciding on the convergence of these series is in fact misleading
because their divergence flows from quite a different source.
}

A few first terms of the expansion of the secular Hamiltonian,
Eq.~\ref{expansion}, are listed below. The free term (for $l=0$) is constant
because it depends on the {\em mean} semi-major axis $a_j$ only. The term with
$l=1$ vanishes identically.

Terms of order $2$ and $3$ may be identified with the quadrupole and octupole
secular Hamiltonian, respectively \citep{Ford2000,Lee2003}:
\begin{eqnarray}
\label{quadropole} 
& & \R^{(i,j)}_2 = \frac{1}{8} \left(3 e_i^2+2\right), \\
\label{octupole}
& & \R^{(i,j)}_3 = -\frac{15}{64}
    \left(3 e_i^2+4\right) e_i e_j \cos({\Delta{\varpi_{i,j}}}).
\end{eqnarray}
Higher order terms are the following:
\begin{eqnarray}
&& \R^{(i,j)}_4 = \frac{9}{1024}
    \Big[ 
    70 \left(e_i^2+2\right) e_i^2 e_j^2 \cos({2\Delta{\varpi_{i,j}}})+\\
    && \quad +\left(15 e_i^4+40 e_i^2+8\right) \left(3 e_j^2+2\right)
    \Big],\nonumber \\
&& \R^{(i,j)}_5 = -\frac{105}{4096}
    \Big[
    7 \left(3 e_i^2+8\right)
    e_j^3 e_i^3 \cos({3\Delta{\varpi_{i,j}}}) +\\
    && \quad +2 \left(5
    \left(e_i^2+4\right) e_i^2+8\right) \left(3
    e_j^2+4\right) e_i e_j \cos({\Delta{\varpi_{i,j}}})
    \Big],\nonumber\\
&& \R^{(i,j)}_6 = \frac{5}{65536} 
    \Big[
    2079 \left(3 e_i^2+10\right) e_i^4 e_j^4
    \cos({4\Delta{\varpi_{i,j}}}) +\\
    && \quad +630 \left(15 e_i^4+80
    e_i^2+48\right) \left(e_j^2+2\right) e_i^2 e_j^2 
    \cos({2\Delta{\varpi_{i,j}}}) +\nonumber\\
    && \quad +10 \left(35 e_i^6+210 e_i^4+168
    e_i^2+16\right) \left(15 e_j^4+40
    e_j^2+8\right)\Big].\nonumber
\end{eqnarray}

We computed the expansion up to the order of 24. This expansion is available on
the request in the form of raw MATHEMATICA input file; also available in
the form of on-line material after publishing this paper.

\subsection{A comparison with the octupole theory}
Here,  we compare the results provided by the secular theory derived in the
previous section with the results obtained with  the help of the  octupole
theory of two planets \citep{Lee2003} which has been obtained through averaging
the perturbation Hamiltonian with the help of von~Zeipel method up to the third
order in $\alpha_{1,2}\equiv \alpha$.  It has been applied to study qualitative
features of the secular dynamics in hierarchical planetary systems (i.e. with
small $\alpha$). A similar theory has been developed by \cite{Ford2000} who
investigated secular dynamics in  hierarchical triple stellar systems with large
separation of the third body. 

To make our discussion  more transparent, we use, in this section, the notation
of \cite{Lee2003}.  Their equations~(13), (14) and (15) have the following form:
\begin{eqnarray}
&& L_1 = \frac{m_0 m_1}{m_0 + m_1} \sqrt{k^2 (m_0 + m_1) a_1},\\
&& L_2 = \frac{(m_0 + m_1) m_2}{m_0 + m_1 + m_2} \sqrt{k^2 (m_0 + m_1 + m_2) a_2},\\
&& G_j = L_j \sqrt{1 - e_j^2},
\end{eqnarray}
where $m_{1,2}$ are planetary masses, $m_0$ is the mass of the star, $j=1,2$
(we set the gravitational constant to $k^2$).  To derive the octupole theory in
terms of Jacobi reference frame, we start from writing down Eq.~\ref{Hfull} with
respect to Jacobi coordinates $\vec{r}_{1,2}$ of two point masses, as a sum of
two Keplerian terms and the perturbation:
\[
{\cal H}^J = \frac{1}{2\mu_1} \vec{p}^2_1
  -\frac{k^2 m_0 m_1}{\|\vec{r}_1\|}
  + \frac{1}{2\mu_2} \vec{p}_2^2 
    -\frac{k^2 m_0 m_2}{\|\vec{r}_2\|} + {{\HpertJ}},
\]
where
\[
{\HpertJ} = 
    -\frac{k^2 m_1 m_2}{\|\vec{r}_2 - (1-\kappa_1) \vec{r}_1\|}
    + k^2 m_0 m_2\left[ \frac{1}{\|\vec{r}_2\|} - 
    \frac{1}{\|\vec{r}_2 + \kappa_1 \vec{r}_1\|}
    \right].
\]
Here, $\vec{p}_{1,2}$ are the conjugate Jacobi momenta,
$\kappa_1=m_1/(m_0+m_1)$, and the reduced
masses are
\[
\mu_1 = \frac{m_1 m_0}{(m_0+m_1)},
 \quad \mu_2=\frac{m_2 (m_0+m_1)}{(m_0+m_1+m_2)}.
\]
For details, see, e.g., \citep{Malhotra1993}. Now,  after expanding ${\HpertJ}$
with respect to small $\kappa_{1}$ and retaining first order terms, we can
show that the  perturbation  has the same form as Eq.~\ref{Hpert} with the
accuracy to the second order in the masses $m_{1,2}/m_0$
\citep{Malhotra1993}:
\[
\Hint = - k^2 m_1 m_2 \left[ \frac{1}{\|\vec{r}_2-\vec{r}_1\|}
 - \frac{\vec{r}_1 \cdot \vec{r}_2}{\|\vec{r}_2\|^3} \right].
\] 
The same truncated Hamiltonian is analyzed in \citep{Libert2005} who derived the
secular Hamiltonian of the 12-th order in eccentricities by the "averaging with
scissors" (i.e., by eliminating  from the Fourier expansion of $\Hint$ all fast
periodic terms dependent on $l_i$). These authors report that their secular
theory reproduces qualitatively results of \cite{Michtchenko2004} on the
analytical way. 

The indirect part of the truncated Hamiltonian $\Hint$ also averages out to a
constant, hence the rest of the averaging process is the same as  in the case of
$\Hpert$ written with respect to  the Poincar\'e elements. Nevertheless, the
averaged $\Hint$ is missing  terms of orders higher than two in the planetary 
masses. Indeed, with our method we derived the secular octupole Hamiltonian
which has the same functional form as formulae (17) in \citep{Lee2003}. 
However, there are some differences in coefficients $C_2$ and $C_3$ [see their
equations (18) and (19)].  These coefficients in our expansion are the
following:
\begin{eqnarray}
&& C_2 = \frac{1}{16} \frac{G^2 (m_0 + m_1)^7 m_2^7}{(m_0 + m_1 + m_2)^3 (m_0 m_1)^3}
\frac{L_1^4}{L_2^3 G_2^3} D_2,\\
&& C_3 = \frac{15}{64} \frac{G^2 (m_0 + m_1)^9 m_2^9 (m_0 - m_1)}{(m_0 + m_1 + m_2)^4 (m_0
m_1)^5} \frac{L_1^6}{L_2^3 G_2^5} D_3,
\end{eqnarray}
where we extracted out two factors leading to the  difference between the
respective formulae:
\begin{eqnarray}
&& D_2 = \frac{m_0 + m_1}{m_0} \sim 1 + O\left(\frac{m_1}{m_0}\right), \\
&& D_3 = \frac{(m_0 + m_1)^2}{m_0 (m_0 - m_1)} \sim 1 +
O\left(\frac{m_1}{m_0}\right).
\end{eqnarray}
These factors can be thought as equal to $1$ in \citep{Lee2003}. However, the
theories are consistent within the assumed accuracy of the expansion and  the
relative magnitude of terms skipped from ${\HpertJ}$ [of the order of
$O(m_1/m_0)$].

Actually, our averaging algorithm can be applied also to the {\em full}
perturbing Hamiltonian, thanks to straightforward generalization for  terms like
the following: 
\[
\frac{1}{\| \delta_1 \vec{r}_1 - \delta_2 \vec{r}_2\|},
\]
where $\delta_1,\delta_2$ are some constants. In that instance, we obtain exactly the
same $C_{2,3}$ as in \citep{Lee2003}. Hence,  as one would expect, both
approaches lead to fully equivalent results.

This comparison also reveals  that the averaging of  the truncated Hamiltonian
is in fact quite problematic  because the accuracy of the secular expansion has
nothing to do with the magnitude of the rejected terms. Already for Jupiter-mass
planets, a contribution of these terms may be significant (see Sect.~2.4 for
details).

Moreover, a direct comparison of the theories would be  more subtle. Although
the functional forms of the perturbing Hamiltonians are the same,  they are
expressed in terms of two different sets of canonical variables. Hence, the mean
elements $a, e, \varpi$ have different meaning in these theories, i.e., the same
physical configuration of the planets will be parameterized with quantitatively
different values of the mean elements.

\subsection{Tests of the analytic secular theory}
\label{sec:tests}
\begin{figure*}
\centerline{
\includegraphics [width=17.6cm]{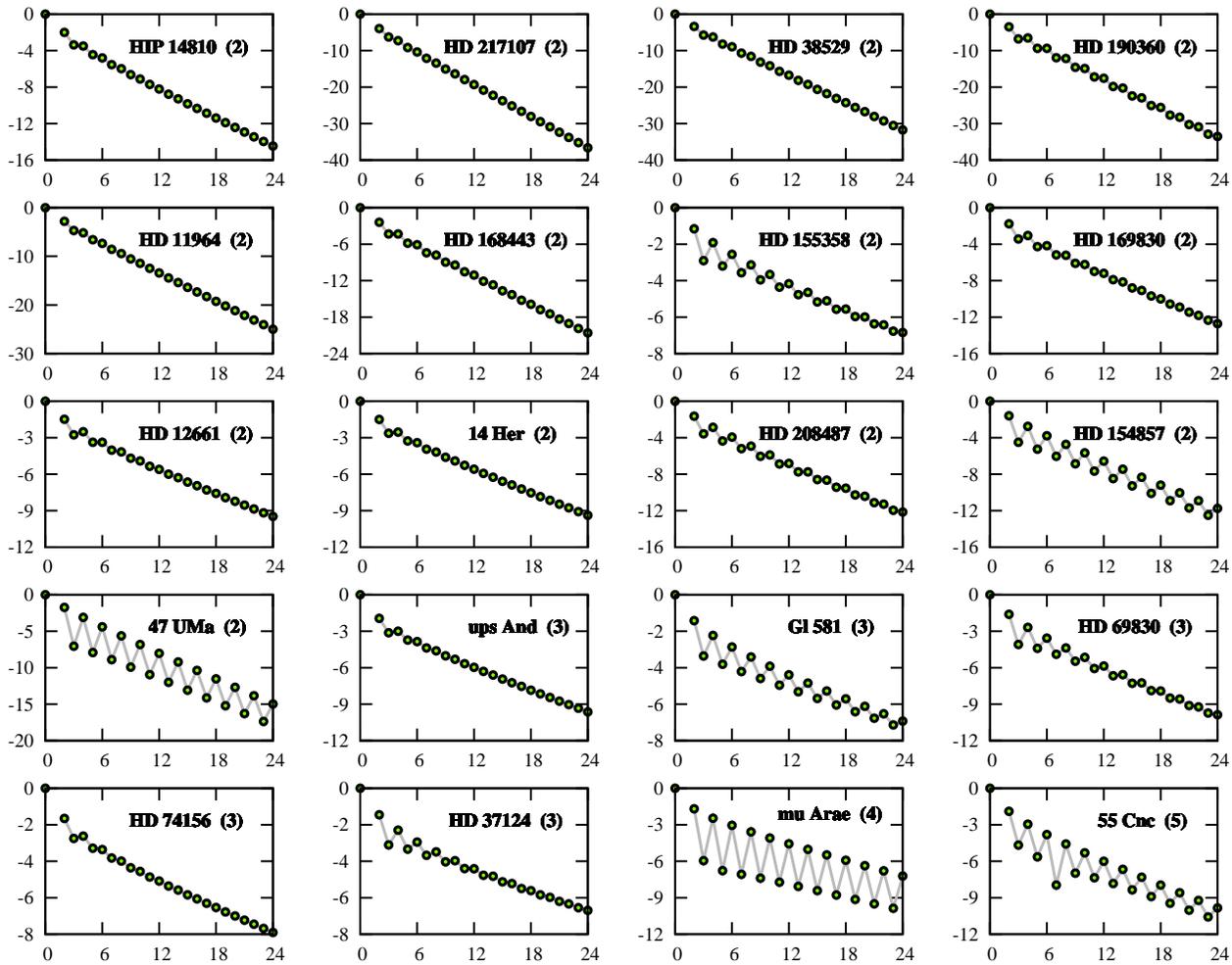}
}
\caption{
Convergence of the expansion of the $N$-planet secular Hamiltonian derived in
this paper.  Each panel is for the magnitude of expansion terms
in~Eq.~\ref{expansion} divided by free term. The $x$-axis is for the order  of
the expansion, the $y$-axis is for  $\log_{10}\,\|{\cal H}_n\|$, where ${\cal
H}_n$ denotes the magnitude of the sum of $n$-th order terms divided by the  sum
of the free terms.  Parameters $(\mu_{i,j},\alpha_{i,j})$ of $\Hsec$  are chosen
for the known multi-planet extrasolar planetary systems. Their parent stars are
marked in each panel together with number of planets written in brackets.
}
\label{accuracy1}
\end{figure*}

To test the accuracy and relevance of the high order expansion of $\Hsec$, we
calculated the  magnitude of  subsequent terms  relative to the free term. This
expansion is computed up to the order of 24 for parameters
($\mu_{i,j},\alpha_{i,j}$) of extrasolar planetary systems taken from  the Jean
Schneider  Encyclopedia of Extrasolar Planets\footnote{http://exoplanets.eu}. 
The results of this experiment are presented in Fig.~\ref{accuracy1}. Each panel
in this figure is labeled with the name of a relevant star and a number of
putative planets it hosts (written in brackets).  As we can see, for all 
examined systems, the sum of terms in $\Hsec$  of the same order (note that we
can have two and more planets in the system) decrease rapidly with the order of
the expansion. For a few most separated systems  with two planets (e.g.,
HD~217107, HD~190360), the highest order terms are as low as $\sim 10^{-37}$ in 
the relative magnitude.  In the tested sample, the largest 24th-order terms are
$\sim 10^{-7}$.  Hence, the secular energy can be  calculated with excellent
accuracy.  We note that similar tests of the precision of the secular theory were
presented in \citep{Gallardo2005} and \citep{Libert2005}.

In the second test,  we compare the outcome of our algorithm with the results of
semi-analytical approach  by \cite{Michtchenko2004,Michtchenko2006} [see also
\cite{Migaszewski2008} for some technical aspects] who applied the method to
study the secular dynamics of two-planet system. In the algorithm of
\cite{Michtchenko2004} the perturbing Hamiltonian is averaged out by means of
the numerical integration. Hence, one avoids any expansion of the Hamiltonian
and the results are formally exact (or very accurate, providing that precise
enough quadratures are applied). 

To examine the  accuracy of the analytic secular theory, we calculated the
levels of the secular Hamiltonians for a number of extrasolar planetary systems
which can be potentially regarded as non-resonant or well fitting assumptions of
the secular theory. Because the phase space of the secular problem is
three-dimensional [i.e., $(G_1,G_2,\Delta\varpi)$], we choose the so called
representative plane of the initial conditions to plot the  secular energy
levels. The definition of the representative plane follows
\cite{Michtchenko2004}. The secular Hamiltonian of a coplanar two-planet system
depends only on $\Delta\varpi=\varpi_2-\varpi_1$ and eccentricities $(e_1,e_2)$
coupled through the integral of the total angular momentum   (or  the so called 
Angular Momentum Deficit, $AMD = G_1 + G_2$). Effectively, the secular problem
has one degree of freedom and is integrable.  \cite{Michtchenko2004} have shown
that {\em all} phase trajectories of the non-resonant system pass through a
plane defined with $\Delta\varpi = 0$ or $\Delta\varpi = \pi$ [see also
\citep{Pauwels1983} for the qualitative analysis of the secular two-planet
problem]. The representative plane may be defined for fixed $\alpha_{1,2} \equiv
\alpha = a_1/a_2$ and $\mu_{1,2} \equiv\mu=m_1/m_2$ as follows:
\[
{\cal S} = \{e_1 \cos\Delta\varpi  \times e_2\, ;\ 
e_1 \in [0,1),\, e_2 \in [0,1), \, \Delta\varpi=0 \, \cup \, \Delta\varpi = \pi
\}.
\]
This plane comprises of two $(x=e_1 \cos\Delta\varpi, y=e_2)$-half-planes with
$x \leq 0$ for $\Delta\varpi = \pi$ and with  $ x \geq 0$ for $\Delta\varpi =0$.
Simultaneously, the derivatives of the secular Hamiltonian with respect to
$\Delta\varpi$ are equal to zero for $\Delta\varpi=0,\pi$.   It follows from the
symmetry of interacting orbits with respect to both apsidal lines.  Having  the
secular Hamiltonian in explicit analytic form, we can verify this property 
directly.  Indeed, each term in the secular Hamiltonian depends on
$\Delta\varpi$ only through $\cos(l \Delta\varpi)$ with $l \in {\mathbb N}$,
$l>0$ and it means that $\Hsec$ is even function of $\Delta\varpi$. Hence,
$\partial\, \Hsec/\partial\,\Delta\varpi$ implies factors involving $\sin(l
\Delta\varpi)$ and these terms vanish identically for  $\Delta\varpi = 0,\pi$. 
{ Formally, it is  possible  that $\partial\, \Hsec/\partial\,\Delta\varpi=0$
also for $\Delta\varpi \neq 0,\pi$, however, to find such solutions we should
solve highly nonlinear equation involving $(e_1,e_2)$ and trigonometric
functions of $\Delta\varpi$.}

Each pair of $(e_1, e_2)$ for which $\partial\, \Hsec/\partial\,G_1=0$
corresponds to an equilibrium in the secular problem (simultaneously, they are
the extrema of the secular Hamiltonian).  These equilibria appear both in the
negative half-plane of ${\cal S}$, as mode~II solutions \rev{(this mode is
Lyapunov stable,  and may be characterized with librations of angle 
$\Delta\varpi$ around $\pi$ in the  evolution of neighboring orbits}), and in
the positive half-plane of ${\cal S}$ as mode~I solutions 
\rev{(Lyapunov stable, with librations of angle $\Delta\varpi$ around $0$
of the nearby orbits).
} 
Further, in the regime of large eccentricities, in the positive half-plane a
new, non-classic mode of motion may appear [it is the so called non-linear
secular resonance,  NSR from hereafter, see \citep{Michtchenko2004} for
details]. These results are derived through the numerical (exact) approach,
hence their reproduction by the analytical theory provides an absolute test of
its quality and accuracy.

\begin{figure*}
\centerline{
\vbox{
    \hbox{\includegraphics [width=58mm]{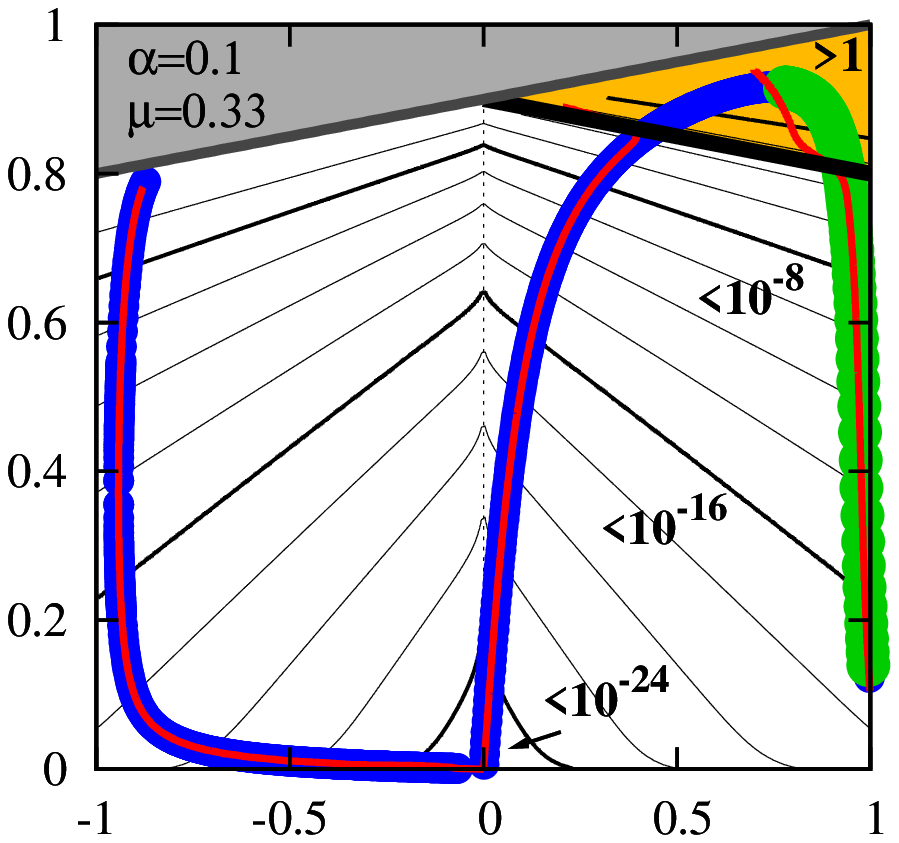}\hskip-3mm
          \includegraphics [width=58mm]{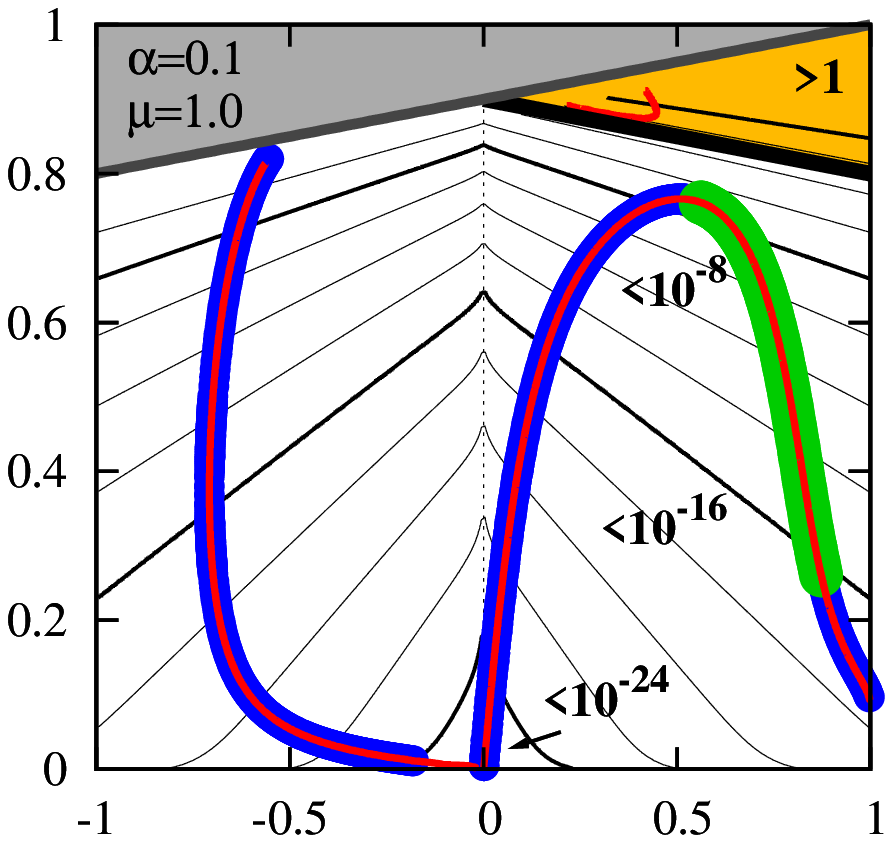}\hskip-3mm
          \includegraphics [width=58mm]{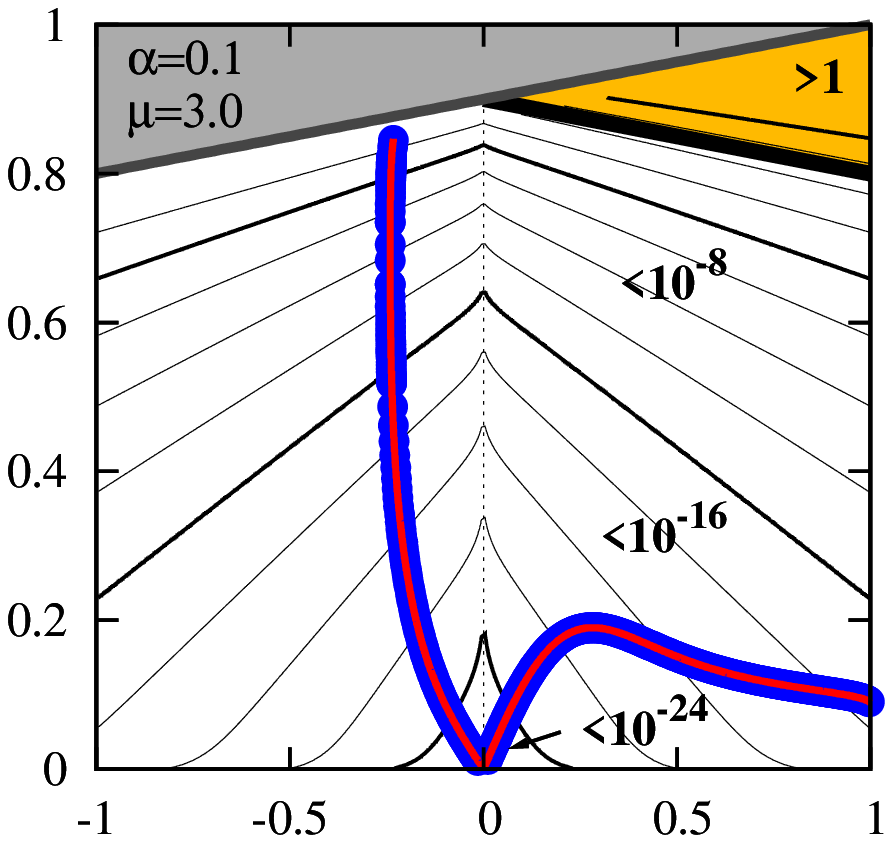}}
    \hbox{\includegraphics [width=58mm]{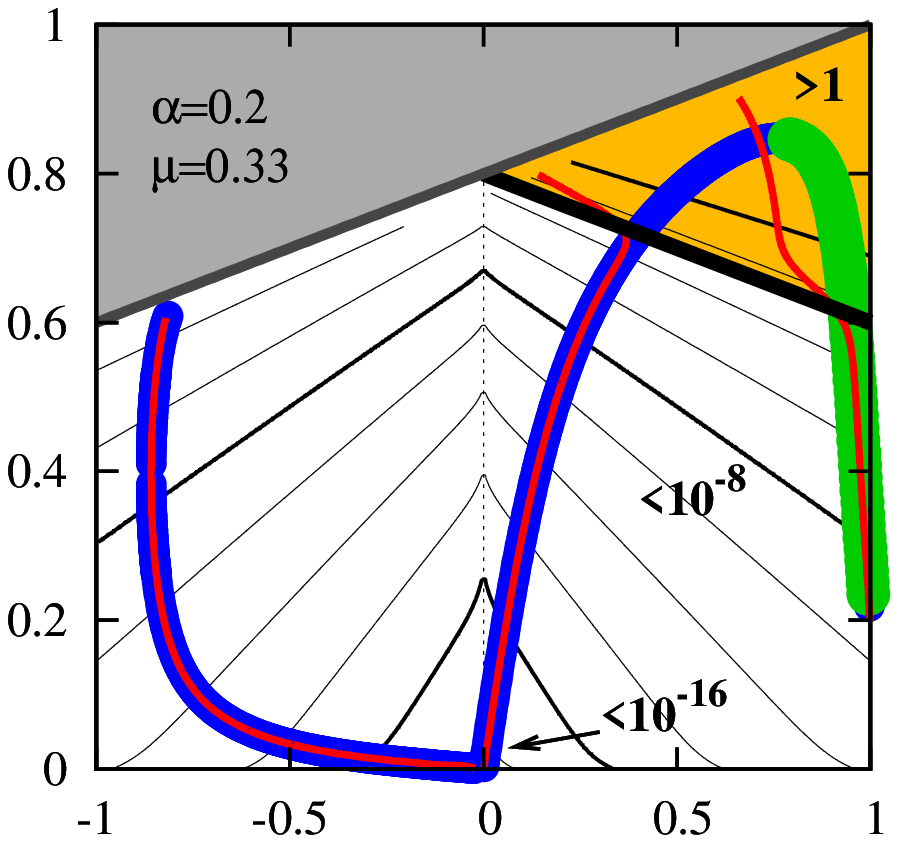}\hskip-3mm
          \includegraphics [width=58mm]{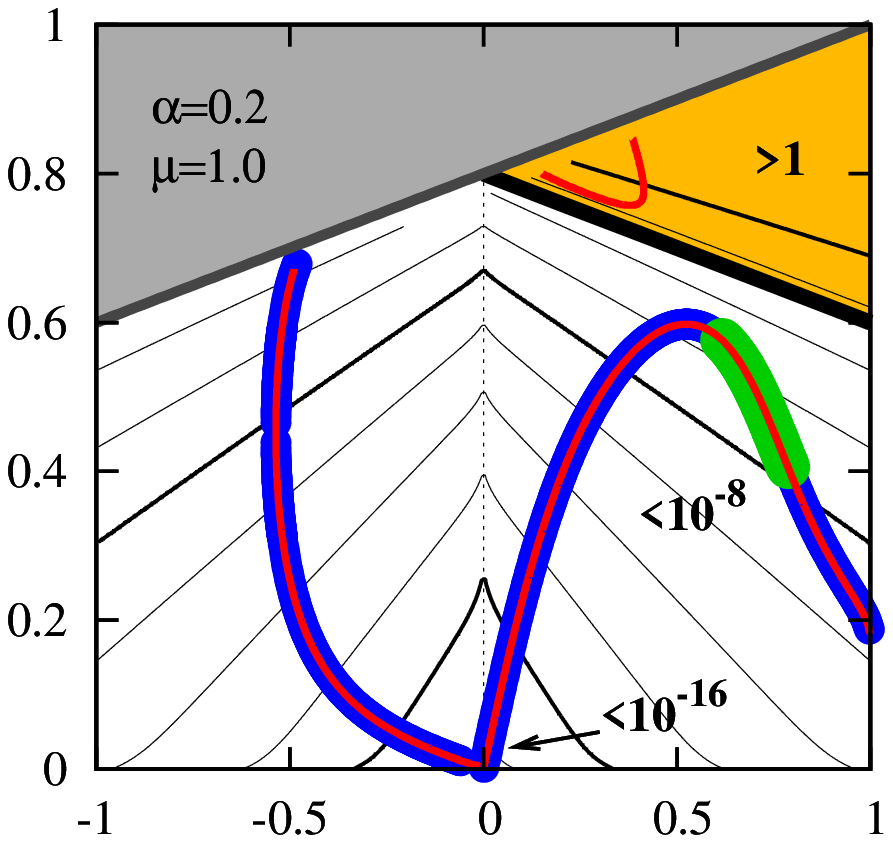}\hskip-3mm
          \includegraphics [width=58mm]{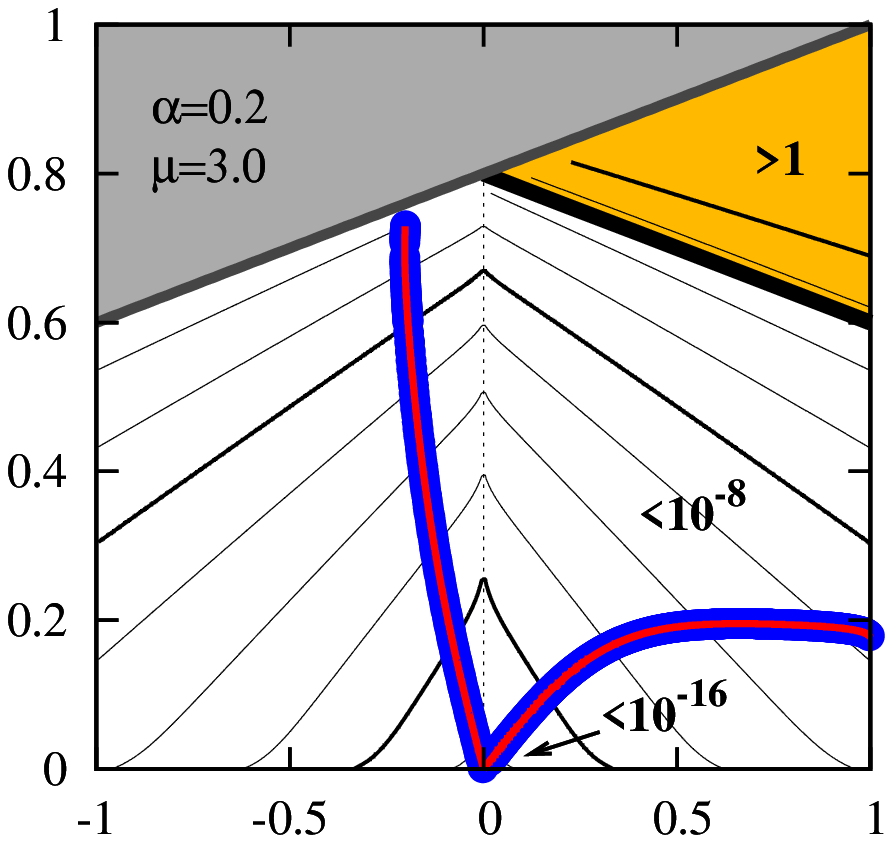}}
    \hbox{\includegraphics [width=58mm]{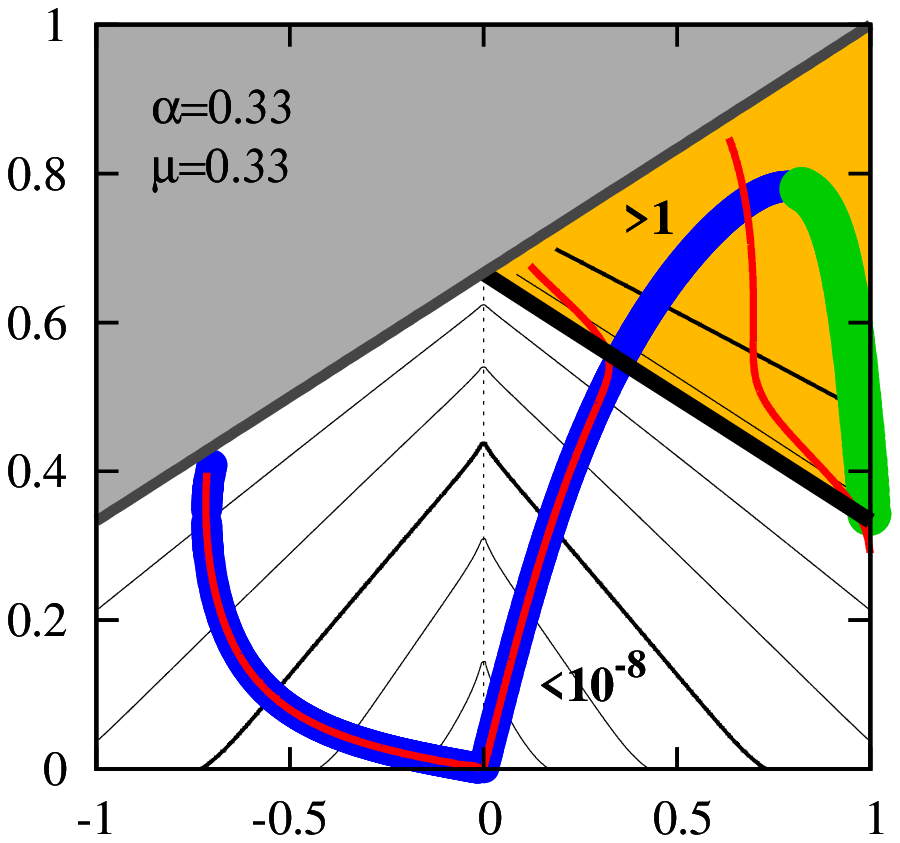}\hskip-3mm
          \includegraphics [width=58mm]{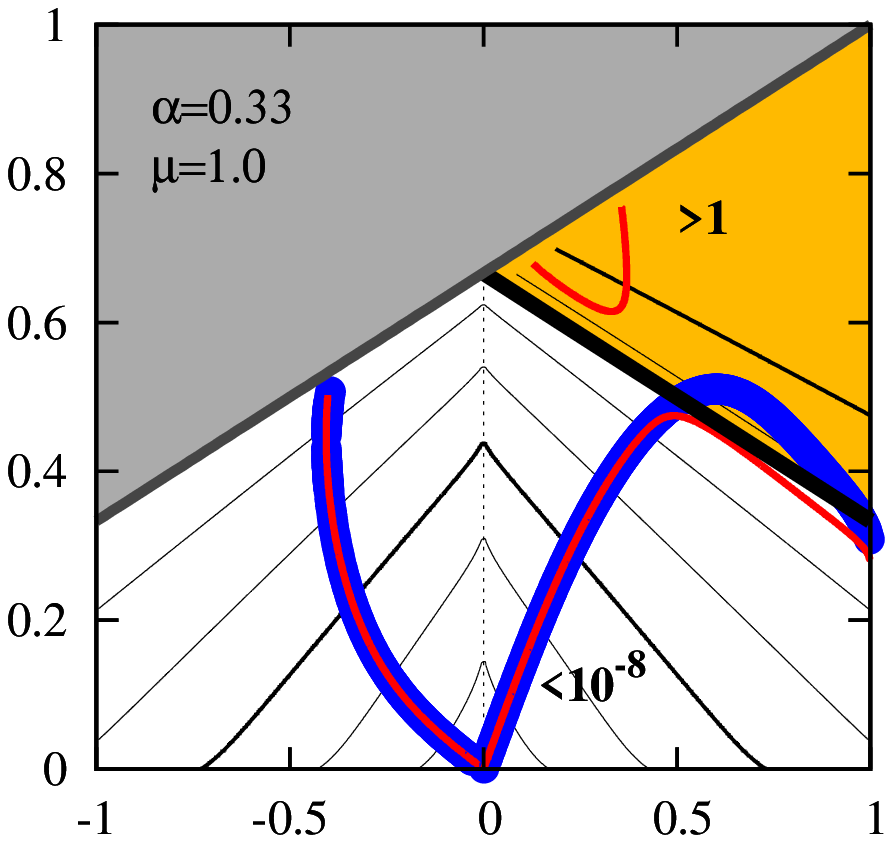}\hskip-3mm
          \includegraphics [width=58mm]{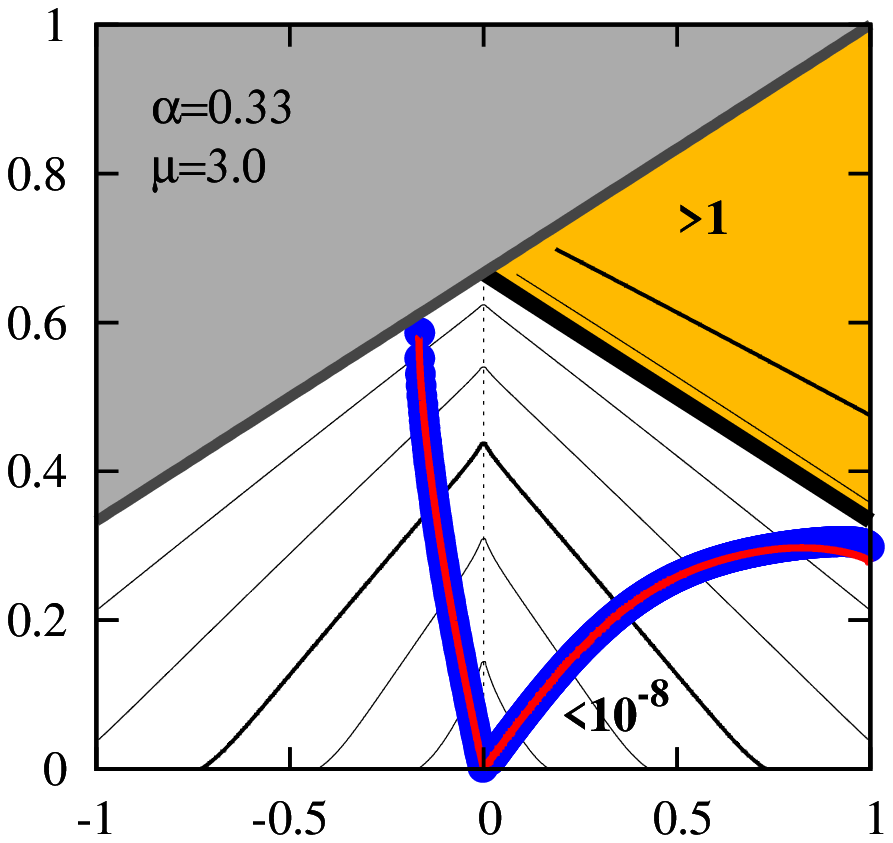}}
}
}
 \caption{
A test of the secular theory of a coplanar two-planet system, derived in this
paper. Each panel is for the representative energy plane,   $(e_1
\cos{\Delta{\varpi}}, e_2)$.  In the right half-plane, $\Delta{\varpi}=0$,  at
the left half-plane, $\Delta{\varpi}=\pi$.   Gray region corresponds to crossing
orbits and its boundary is defined through $a_2 (1 - e_2) = a_1 (1 -  e_1)$,
where (1) is for the inner orbit and (2) is for the outer orbit. Black
thick line marks the ``anti-collision'' line defined with $a_2 (1 - e_2) = a_1
(1 + e_1)$ (see the text for more details). Black, thin lines are for  contour levels of  the last two terms
of expansion Eq.~\ref{expansion}, i.e., the sum of absolute values of the
$23$-th and $24$-th order terms divided by the free term.   A few contour
levels  ($1$, $10^{-8}$, $10^{-16}$ and $10^{-24}$, respectively) are  marked
with  thicker curves. The area colored in orange determines the region where the
expansion of $\Hsec$ is divergent. Thick blue  lines mark the positions of
stable equilibria: mode~II (with apsides anti-aligned, the left half-plane of
the representative plane), and mode~I (with apsides  aligned, the right-half
plane).  These solutions are obtained numerically with the help of
semi-analytical averaging algorithm in \citep{Michtchenko2004}. Green curves are
for the nonlinear-secular resonance (NSR).  Red curves are for the corresponding
libration modes  calculated analytically with the help of the 24-order expansion
of $\Hsec$. Labels in the top-left corner at each panel are for the parameters
of the expansion, $\alpha \equiv a_1/a_2$  and $\mu \equiv m_1/m_2$.
}
\label{accuracy2}	  
\end{figure*}
The results of  the second experiment are presented in Figure~\ref{accuracy2}. 
Each panel in this figure is for the representative plane of initial conditions
computed and calculated for two-planet systems characterized with mass ratio
$\mu$ and semi-major axes ratio $\alpha$. These parameters are  written in the
top-left corner at each relevant panel. To illustrate the significance of
highest order terms in Eq.~\ref{expansion}, we plot contour levels of  
$(\|{\cal H}_{23}\| + \|{\cal H}_{24}\|)/\|{\cal H}_0\|$, i.e.,  the relative
magnitude of the sum of the last two terms with respect to the magnitude of the
free term (in general, ${\cal H}_n$ would stand for a sum of expansion terms
over the number of planet pairs in the given multi-body configuration). Four
particular contour levels of $1$, $10^{-8}$, $10^{-16}$  and $10^{-24}$,
respectively, are distinguished with thicker lines and labeled accordingly.

In each panel, we also marked positions of stationary solutions (modes I, II,
and~NSR).  Solutions represented by  thick curves are derived with the help of
numerical averaging \citep{Michtchenko2004}: blue curves are for stable
equilibria, and  thick green curves are for unstable solutions (the {\em
non-linear secular resonance}). These curves represent practically exact (or
very precise) solution to the problem. The thin, red lines mark the positions of
equilibria calculated analytically, with the help of the $24$th-order expansion
of $\Hsec$.

The results can be summarized with a few  interesting conclusions. Clearly, in
the regions of the representative  $(e_1\cos\Delta\varpi,e_2)$-plane at which 
$(\|{\cal H}_{23}\| + \|{\cal H}_{24}\|)/\|{\cal H}_0\| < 10^{-3}$, the
precision of the analytical method is excellent. The secular theory predicts
exactly positions of the equilibria in the negative half-plane of ${\cal S}$
(when $\Delta{\varpi} = \pi$) no matter how large is $\alpha$. On contrary, the
exact derivation of the shape of the non-linear secular resonance is a
challenging problem for the analytic approach \citep[see also][]{Henrard2005}.
The nonlinear resonance can be reproduced well by the analytic theory providing
that $(\|{\cal H}_{23}\| + \|{\cal H}_{24}\|)/\|{\cal H}_0\| < 10^{-3}$. This is
also an empirical border of the convergence of the secular expansion.  We also
found another empirical convergence condition that follows from the notion of
the geometric series, i.e,
$
\|{\cal H}_{24}/{\cal H}_{22}\|<1.
$
This inequality is illustrated with triangular, orange colored regions labeled
with ``$>1$''. In these regions, the series are divergent, hence the secular
theory cannot reproduce the {\em real} dynamics.  This may be interpreted as a
clear limitation of the analytic theory. In fact, as we show below (Sect.~2.5),
this problem appears rather due to imperfect algorithm of the expansion,  which
can be still improved. On the other hand, the results of this test  provide an
example that illustrates excellent properties of the semi-analytical  approach
invented by \cite{Michtchenko2004}.

{
\subsection{An improved averaging algorithm}
A real source of the divergence of the secular series, Eq.~\ref{expansion}, can
be deduced after we draw in Fig.~\ref{accuracy2} the ``anti-collision'' line 
defined through $a_2 (1 - e_2) = a_1 (1 + e_1)$ (see the right-half of the
representative plane). Clearly, the series diverge above this line  and the
positions of the equilibria are strongly distorted. In this area, for some
points or parts of the orbits, $r_i[E_i(t)]>r_j[f_j(t)]$, while in the expansion
Eq.~\ref{expansion},  $r_i<r_j$ {\em must} be satisfied in the
whole ranges of the anomalies.  That may  happen when the pericenter of
the outer orbit is closer to the star than the apocenter of the inner orbit, no
matter what is the relative orientation of their
apsidal lines. Then the condition of $r_i<r_j$, which required to write down
Eq.~\ref{eq:squareroot2}, is violated. Obviously, it may be expressed by the
equation of the anti-collision line, and in other
words, through the requirement that the inner orbit lies inside a circle of a radius equal to
the pericenter distance of the outer orbit. Now it is also clear why for the
apsides anti-aligned, we have always very good convergence of the secular series
while in the right half-plane, the convergence region is generally 
strongly limited. Hence, the convergence limit of the expansion described in Sect.~2 may
be simply interpreted through the conditions for the collision lines.
We may also note that the problem persists in any secular theory that relies
directly on Eq.~\ref{eq:squareroot2}. One should be also aware that the
conditions of crossing orbits do not depend on masses.
If the masses are large, the  dynamical collision curve
appears for much smaller eccentricity than the geometrical collision line.
\citep[e.g.,][]{Gozdziewski2007,Michtchenko2008}.

A cure for the divergence problem  may be a modified expansion of the term
$1/\Delta_{i,j}$, helping us to construct  a secular  theory that has no limits
of the type $a_i (1 + e_i) < a_j (1 - e_j)$. To derive the secular  expansion in
Sect.~2, at first we factor $r_j$ in Eqs. (\ref{squareroot1}), 
(\ref{eq:squareroot2}). To have the series convergent for all positions of
planets on their orbits, we propose to factor from the square
root~(\ref{squareroot1}) a term consisting of $a_2$ multiplied by some scale
factor, $\eta \geq 1$ (instead of $r_j$). Then we can express the distance
between planets $i$ and $j$ as follows:
\begin{equation}
\Delta_{i,j} = \eta a_j \sqrt{1 + \zeta_{i,j} (\vec{r}_i,\vec{r}_j)},
\label{newexp}
\end{equation}
where
\begin{equation}
\zeta_{i,j} = \frac{1}{\eta^2 a_j^2} \left[r_i^2 + r_j^2 - 2 \vec{r}_i \cdot \vec{r}_j - \eta^2
a_j^2\right] \equiv 
 \frac{1}{\eta^2 a_j^2} \left[\Delta_{i,j}^2 - \eta^2 a_j^2\right].
\label{zeta}
\end{equation}
The requirement of the convergent expansion of $\Delta_{i,j}^{-1}$ with respect
to $\zeta_{i,j}$  implies $\|{\zeta}_{i,j}\|<1$. If $\|\zeta_{i,j}\|=1$  the
distance between planets is equal to $\sqrt{2} \eta a_j$. The convergence
condition is fulfilled when   $\max \Delta_{i,j} < \sqrt{2} \eta a_j$ for the
given orbits. The maximal distance between planets in coplanar orbits may be
bounded by $a_i (1 + e_i) + a_j (1 + e_j)$. Then  the factor $\eta$ has the 
form of:
\begin{equation}
\eta = \frac{a_i (1 + e_i) + a_j (1 + e_j)}{\sqrt{2} a_j} = \frac{1}{\sqrt{2}}
\left[\alpha_{i,j} (1 + e_i) + (1 + e_j)\right].
\label{eta}
\end{equation}
For hierarchical systems with small eccentricity of the outer planet, $\eta
\approx 1$. For more compact systems with  large eccentricities, $\eta \approx
\sqrt{2}$. For non-coplanar systems, $\eta$ may be  even larger. In
practice, this parameter  makes it possible to control the convergence rate of
the secular expansion. The convergence rate will be faster for large distances
$\Delta$ but slower for smaller distances, moreover the condition of
$-1<\zeta_{i,j}<1$ should be always fulfilled.

The term $1/\Delta_{i,j}$ may be expanded with respect to $\zeta_{i,j}$:
\begin{equation}
\frac{1}{\Delta_{i,j}} = \frac{1}{\eta a_j}
\left[ 1 +  \sum_{l=1}^{\infty} {\frac{(-1)^l (2 l - 1)!!}{2^l l!} 
\zeta_{i,j}^l} \right].
\label{exp2}
\end{equation}
To average out the above formulae, we express positions of both planets in a
given pair with respect to  the eccentric anomaly. Next, we change the
integration variables similarly as in Sect.~2, i.e.,
$
d\,M_k = I_k(E_k, e_k) d\,E_k.
$ 
We also should express $\vec{r}_i \cdot \vec{r}_j$ through $\Delta{\varpi}$
(again, fixing the reference frame with the apsidal line of the inner orbit).
After expressing terms $r_i, r_j, \vec{r}_i \cdot \vec{r}_j$ through eccentric
anomalies, the  function $\zeta \equiv \zeta_{1,2}$ has the
following explicit form
(to shorten the notation, let us fix $i\equiv 1$, $j \equiv 2$ for a
given pair of planets):
:
\begin{eqnarray}
&&\zeta = \frac{1}{\eta ^2} \Big[\theta_0 + \theta_1 \cos{E_1} + \theta_2 \cos{E_2} +
\theta_3 \sin{E_1} + \\ 
&& + \theta_4 \sin{E_2} +
\theta_5 \cos{E_1}^2 + \theta_6 \cos{E_2}^2 + \theta_7 \cos{E_1} \cos{E_2} + \nonumber\\
&& + \theta_8 \sin{E_1}
\sin{E_2} + \theta_9 \cos{E_1} \sin{E_2} + \theta_{10} \cos{E_2} \sin{E_1}\Big],\nonumber
\end{eqnarray}
where ${E}_1, E_2$ are eccentric anomalies of the inner and outer planets
respectively, and coefficients
$\theta_l, l \geq 0$, read as follows:
\begin{eqnarray}
&& \theta_0 = \alpha^2-2 \alpha e_1 e_2 \cos{\Delta{\varpi}}+1-\eta^2,\nonumber\\
&& \theta_1 = 2 \alpha e_2 \cos{\Delta{\varpi}}-2 \alpha^2 e_1,\nonumber \\
&& \theta_2 = 2 \alpha e_1 \cos{\Delta{\varpi}}-2 e_2,\nonumber \\
&& \theta_3 = -2 \alpha e_2 \sqrt{1-e_1^2} \sin{\Delta{\varpi}},\nonumber \\
&& \theta_4 = 2 \alpha e_1 \sqrt{1-e_2^2} \sin{\Delta{\varpi}},\nonumber \\
&& \theta_5 = \alpha^2 e_1^2, \\
&& \theta_6 = e_2^2, \nonumber \\
&& \theta_7 = -2 \alpha \cos{\Delta{\varpi}}, \nonumber \\
&& \theta_8 = -2 \alpha \sqrt{1-e_1^2} \sqrt{1-e_2^2} \cos{\Delta{\varpi}},\nonumber\\
&& \theta_9 = -2 \alpha \sqrt{1-e_2^2} \sin{\Delta{\varpi}},\nonumber \\
&& \theta_{10} = 2 \alpha \sqrt{1-e_1^2} \sin{\Delta{\varpi}}.\nonumber
\end{eqnarray}
Here, $\alpha \equiv \alpha_{1,2}$, $\Delta{\varpi} \equiv
\Delta{\varpi_{1,2}}$  and $e_1, e_2$ are  the eccentricities of the inner and
outer planet, respectively.

After the double averaging of $\zeta$ over the mean anomalies we 
obtain:
\begin{equation}
<\zeta> = \frac{1}{2 \eta^2} \bigg[\left(3 e_1^2+2\right) \alpha^2-9 \alpha
    e_1 e_2 \cos{\Delta{\varpi}}
    +3 e_2^2-2 \eta^2+2\bigg].\nonumber
\end{equation}
The averaging of the square of $\zeta$ over  the mean anomalies brings the
following formulae:
\begin{eqnarray}
&&<\zeta^2> = \frac{1}{8 \eta^4} 
    \bigg[
    \alpha^4 \left(8 + 40 e_1^2 + 15 e_1^4\right) + \nonumber\\
&&  + \alpha^3 \left(-30 e_1 e_2 (3 e_1^2 + 4) \cos{\Delta{\varpi}}\right) + \\
&&  + \alpha^2 \left[(2 - \eta^2) (16 + 24 e_1^2) + e_1^2 e_2^2 (72 + 100 \cos{2\Delta{\varpi}})
+ 48 e_2^2 \right] + \nonumber\\
&& + \alpha \left(-6 e_1 e_2 (20 - 12 \eta^2 + 15 e_2^2) \cos{\Delta{\varpi}}\right) + \nonumber\\
&& + 8 (\eta^2 - 1)^2 + 40 e_2^2 - 24 e_2^2 \eta^2 + 15 e_2^4
    \bigg].\nonumber
\end{eqnarray}
These preliminary calculations show that the new algorithm leads to more complex
expansion of the secular Hamiltonian than the simple approach in Sect.~2.2
which, as we have demonstrated, is limited in some cases. Moreover, we found
this improvement after submitting the manuscript, hence the new expansion and  a
detailed study of its properties would make the paper very lengthy.   We are
going to present the improved algorithm and the results of its tests in a new
work devoted to the analytic theory of non-coplanar model of $N$-planets. A
generalization of Eq.~\ref{eta} for that case seem straightforward, because we
should only calculate $\vec{r}_i \cdot \vec{r}_j$ with the help of appropriate
rotation matrix parameterized through Euler angles, i.e., the Keplerian elements
($i_i,i_j,\omega_i,\omega_j,\Omega_i,\Omega_j$). Hence, only the coefficients
$\theta_l$ will be modified.
}

\section{Secular dynamics of three-planet system}
The two-planet  secular Hamiltonian in Sect.~2 can be easily adapted to
construct the secular theory for $N$-planet system.  At present, a few
candidates of such configurations are already discovered, including four planet
systems, e.g., $\mu$~Arae
\citep{Jones2002,Butler2006,Gozdziewski2007a,Pepe2007},  five or even six planet
configuration around 55~Cnc \citep{Fischer2007}.  Here, as the simplest and
most  natural generalization of the two-planet model, we consider the secular
theory of $three$-planet configuration which is far from MMRs
and collision zones.

The three-planet model is described by Hamiltonian in Eqs.~(\ref{Hfull}),
(\ref{HKepl}) and (\ref{Hpert}), respectively, where $N=3$.  Because the nodal
longitudes are undefined in the coplanar system, and the secular dynamics
depends on the relative positions of the {\em mean} orbits, we can eliminate the
nodal longitudes from the problem. Let indices $i=1,2,3$ enumerate the planets.
Their semi-major axes are $a_1 < a_2 < a_3$, respectively. After averaging
$\Hfull$ over the mean anomalies, the secular Hamiltonian $\Hsec$ does not
depend on $l_i$ anymore. Therefore, conjugate momenta $L_i$ (hence,  the
semi-major axes) are constants of motion. Because the secular system does not
depend on  particular longitudes of nodes, the respective degree of freedom is
also irrelevant for the secular dynamics.

Hence, the secular  system can be described with the following set of canonical
elements:
\begin{eqnarray}
&& g_1 = -\varpi_1, \qquad G_1,  \nonumber\\
&& g_2 = -\varpi_2, \qquad G_2,          \\
&& g_3 = -\varpi_3, \qquad G_3. \nonumber
\label{trans2}
\end{eqnarray}
We can eliminate one more degree of freedom with the help of the angular
momentum  integral, which can be also expressed with $AMD = G_1 + G_2 + G_3$. 
For that purpose, we perform the following canonical transformation:
\begin{eqnarray}
\label{trans3}
&& \sigma_1 = g_1 - g_3 \equiv \varpi_3 - \varpi_1 \equiv
\Delta\varpi_{1,3},  \quad G_1, \nonumber\\
&& \sigma_2 = g_2 - g_3 \equiv \varpi_3 - \varpi_2 \equiv
\Delta\varpi_{2,3},  \quad G_2, \\
&& \sigma_3 = g_3 \equiv -\varpi_3,  \quad AMD = G_1 + G_2 + G_3, \nonumber
\end{eqnarray}
introducing new canonical angles $\sigma_1,\sigma_2,\sigma_3$.  These angles 
can be interpreted as two-planet $\Delta\varpi$ defined for each pair of planets
in the three-planet system. The secular Hamiltonian can be expressed through
$\sigma_1$ and $\sigma_2$ explicitly, hence $\sigma_3$ is cyclic and $AMD$ is
constant of motion. Actually, we reduced the secular system of three planets to
two degrees of freedom, with the secular energy and $AMD$ as free parameters.

\subsection{Representative planes of initial conditions}
Now, we have a similar problem as in the case of two-planet configuration. We
want to illustrate the dynamical properties of the system in possibly global
manner. To characterize its  dynamical states, we follow the general idea of the
representative plane of initial conditions. Here, we focus on the equilibria in
the secular problem. Fixing the integral of $AMD$ as a parameter of the system,
the dynamics may be represented in four-dimensional phase space of
$(G_1,G_2,\sigma_1,\sigma_2)$, or $(e_1,e_2,\sigma_1,\sigma_2)$. The
representative plane will be chosen according with:
\begin{equation}
\frac{\partial \Hsec}{\partial \sigma_1} = 0, \quad
\quad \frac{\partial \Hsec}{\partial \sigma_2} = 0.
\label{condition_non_axial}
\end{equation}
Then any pair of points $(e^0_1,e^0_2)$ that belongs to
the representative plane, and the following equations
are satisfied:
\begin{equation}
\frac{\partial \Hsec}{\partial G_1} = 0, \quad
\quad \frac{\partial \Hsec}{\partial G_2} = 0,
\label{conditionG}
\end{equation}
defines an equilibrium of the secular problem. Note that the last conditions
implies also $\partial\,\Hsec/\partial\,G_3=0$ because $G_3 \equiv G_3(G_1,G_2)$
is a function parameterized by the total angular momentum (or $AMD$).  The
explicit and simple transformation between the eccentricity and the element~$G$
makes it possible to solve the above conditions in terms of $(e_1,e_2)$.

Actually, the most obvious definition  of the representative plane is  a
generalization of that plane constructed  in the two-planet problem. For fixed
$a_1,a_2,a_3$, $m_1,m_2,m_3$ and $AMD$, as a parameter, the {\em symmetric}
representative plane is the set of points such that
\[
{\cal S} = \{e_1 \cos\sigma_1  \times e_2 \cos\sigma_2 ;\ 
e_1 \in [0,1),\, e_2 \in [0,1), \, \sigma_{1,2}=0 \cup \pi
\}.
\]
This plane comprises of four quaters. The signs of $e_{1,2}$ of the  coordinated
axes tell us on the respective values of the secular angles. In that case, the
condition in Eq.~\ref{condition_non_axial} is fulfilled thanks to apsidal
symmetries of the problem. It is also obvious by recalling that ${\Hsec}$ is
even function of $\sigma_{1,2}$. The derivatives of ${\Hsec}$ over
$\sigma_{1,2}$ depend on  factors involving $\sin (l\sigma_{1,2})$, $l \in
{\mathbb N}$, $l>0$ that vanish for $\sigma_{1,2} = 0,\pi$.  Alternatively, the
representative plane may be also defined through $\sin{\sigma_{1,2}}=0$.

Moreover the condition for the representative plane, defined through  the
vanishing derivatives over secular angles, may be also fulfilled for
$\sin{\sigma_{1,2}} \neq 0$. Let us start with the octupole (third-order)
approximation of the secular Hamiltonian. Using Eqs.~(\ref{expansion}),
(\ref{quadropole}) and (\ref{octupole}),  we can write the secular Hamiltonian
in the following short form of:
\begin{equation}
\Hsec = \gamma_{1,2} \cos{\Delta{\varpi}_{1,2}} + \gamma_{1,3} 
\cos{\Delta{\varpi}_{1,3}} + \gamma_{2,3} 
\cos{\Delta{\varpi}_{2,3}} + \gamma_{4}.
\end{equation}
Using the canonical angles defined with Eq.~\ref{trans3}:
\begin{equation}
\Hsec = \gamma_{1,2} \cos{(\sigma_1-\sigma_2)} + \gamma_{1,3} \cos{\sigma_1} + 
\gamma_{2,3} \cos{\sigma_2} + \gamma_{4},
\end{equation}
where $\gamma_{1,2}, \gamma_{1,3}, \gamma_{2,3}, \gamma_4$ are functions of $e_1, e_2, e_3$.  
Hence, the {\em non-symmetric representative plane} is defined through the following
conditions, the same as Eq.~\ref{condition_non_axial}, 
in the explicit form:
\begin{eqnarray}
-\gamma_{1,2} \sin{(\sigma_1-\sigma_2)} - \gamma_{1,3} \sin{\sigma_1}=0,\nonumber\\
\gamma_{1,2} \sin{(\sigma_1-\sigma_2)}  - \gamma_{2,3} \sin{\sigma_2}=0.
\label{representative_conditions}
\end{eqnarray}
Obviously, conditions in Eq.~\ref{representative_conditions} are satisfied not only
when $\sin{\sigma_{1,2}} = 0$.  We have four other solutions, satisfying 
Eq.~\ref{representative_conditions}, i.e.:
\begin{eqnarray}
\sigma_1 = \pm \arccos \left[-\frac{1}{2} \gamma_{1,2} \left(\frac{1}{\gamma_{2,3}} + 
\frac{\gamma_{2,3}}{\gamma_{1,2}^2} - \frac{\gamma_{2,3}}{\gamma_{1,3}^2}\right)\right],\nonumber\\
\sigma_2 = \pm \arccos \left[-\frac{1}{2} \gamma_{1,3} \left(\frac{1}{\gamma_{1,2}} 
+ \frac{\gamma_{1,2}}{\gamma_{1,3}^2} - \frac{\gamma_{1,2}}{\gamma_{2,3}^2}\right)\right].
\end{eqnarray}
These solutions describe the \textit{non-symmetric} representative planes,  
with respect to the octupole theory. This definition is exact up to the third
order in $\alpha_{i,j}$. Angles $\sigma_1, \sigma_2$ satisfying condition in
Eq.~\ref{condition_non_axial}, i.e., solutions to 
Eq.~\ref{representative_conditions} can be found consistent with higher order
expansions. However,  these equations are very complex and, in practice, we
would have to solve them numerically (for instance, with the Newton-Raphson
algorithm initiated with starting conditions derived from the octupole theory).
Hence, in general, the representation of the energy levels with the help of the
non-symmetric representative planes is much more difficult than in the symmetric
case and is not unique (it has some analogy to the Poincar\'e cross-section).
For instance, we can define the representative plane for higher order expansions
of $\Hsec$. In this paper, we focus on the symmetric representation only.

\subsection{Energy levels for three-planet secular model}
In this section, to show some applications of the secular theory,  we
investigate qualitative dynamics of a few three-planet extrasolar systems. To
characterize  these systems, we calculated  energy levels in the
\textit{symmetric}  representative plane. We also try to find equilibria in the
secular model of each examined system.
\begin{figure*}
 \centerline{
 \vbox{
    \hbox{\includegraphics [width=45mm]{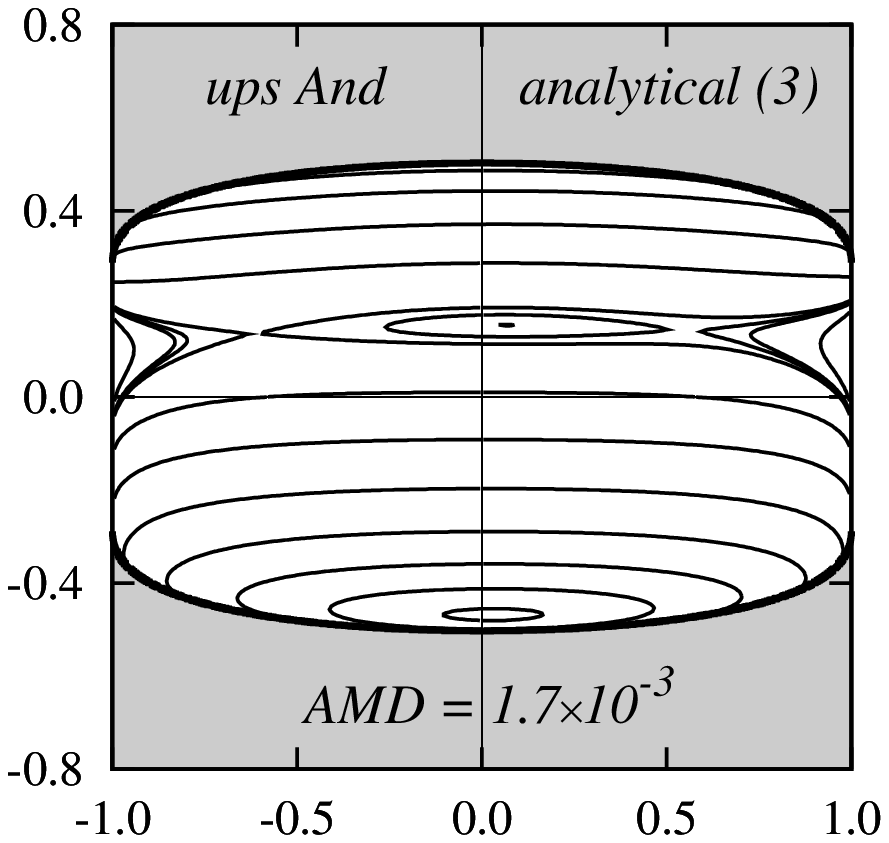}\hskip-3mm
          \includegraphics [width=45mm]{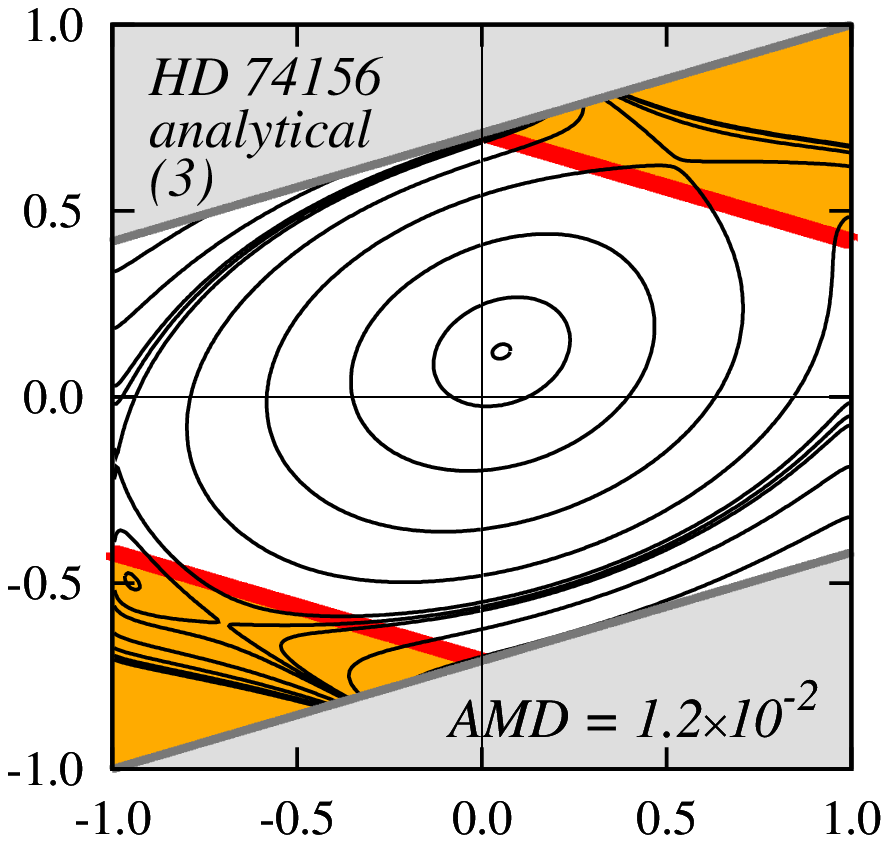}\hskip-3mm
          \includegraphics [width=45mm]{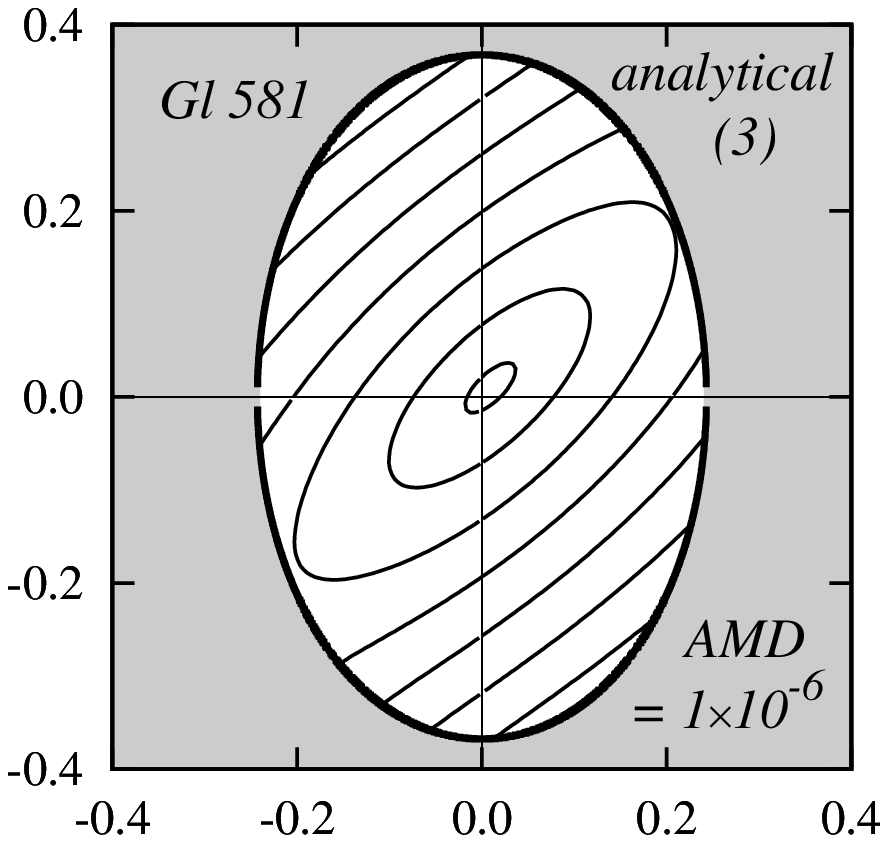}\hskip-3mm
          \includegraphics [width=45mm]{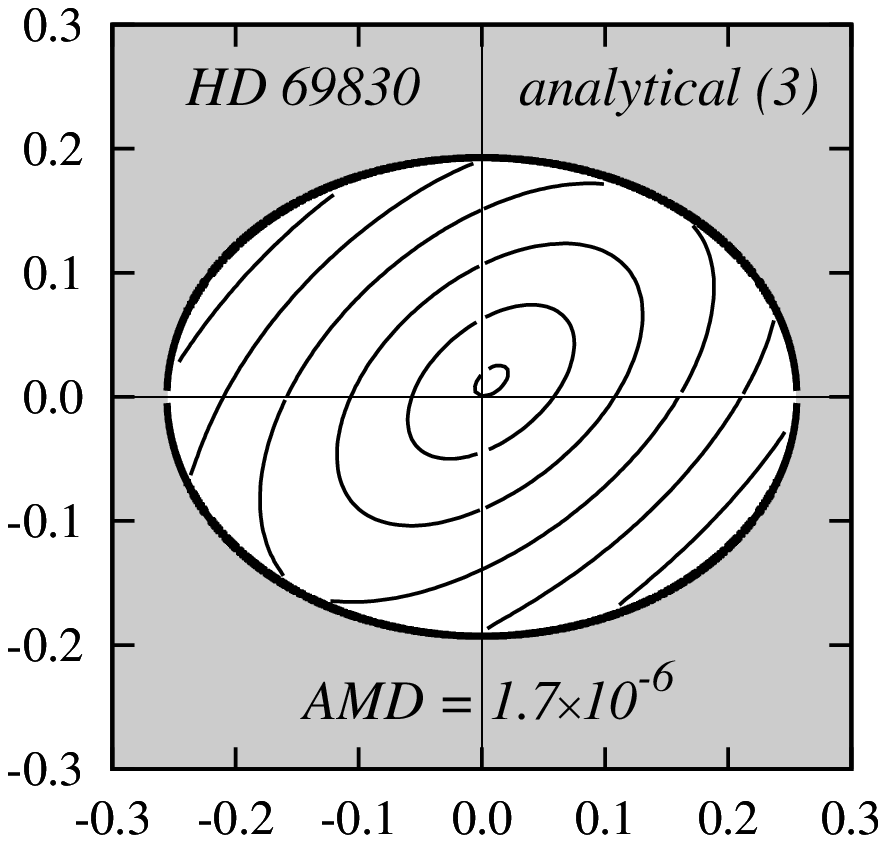}}
    \hbox{\includegraphics [width=45mm]{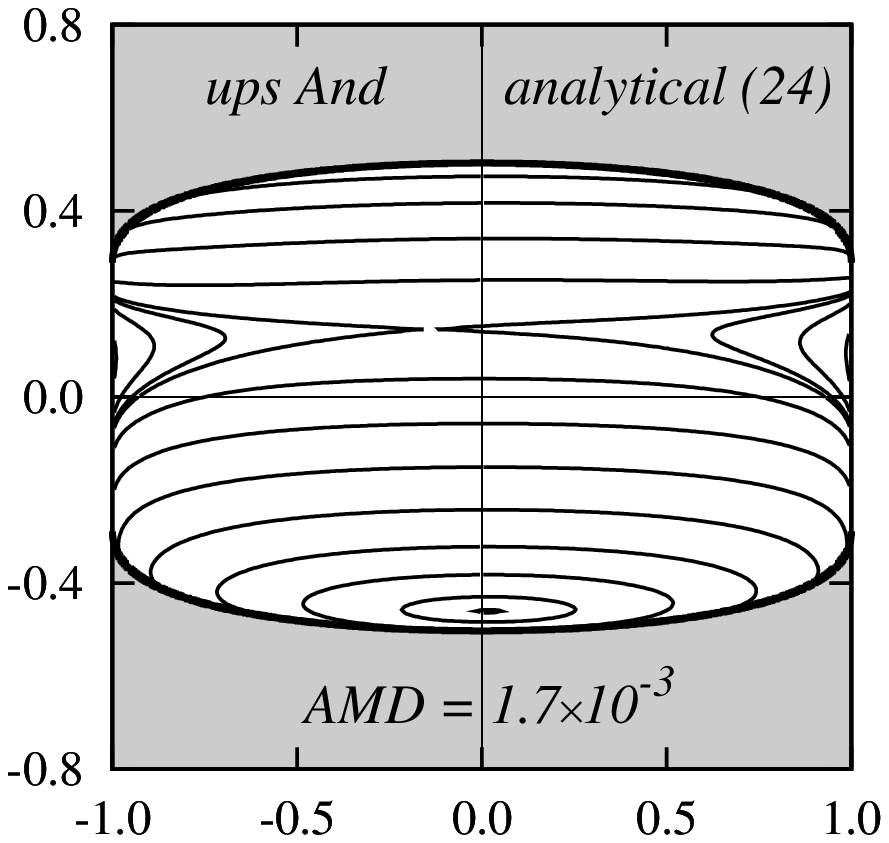}\hskip-3mm
          \includegraphics [width=45mm]{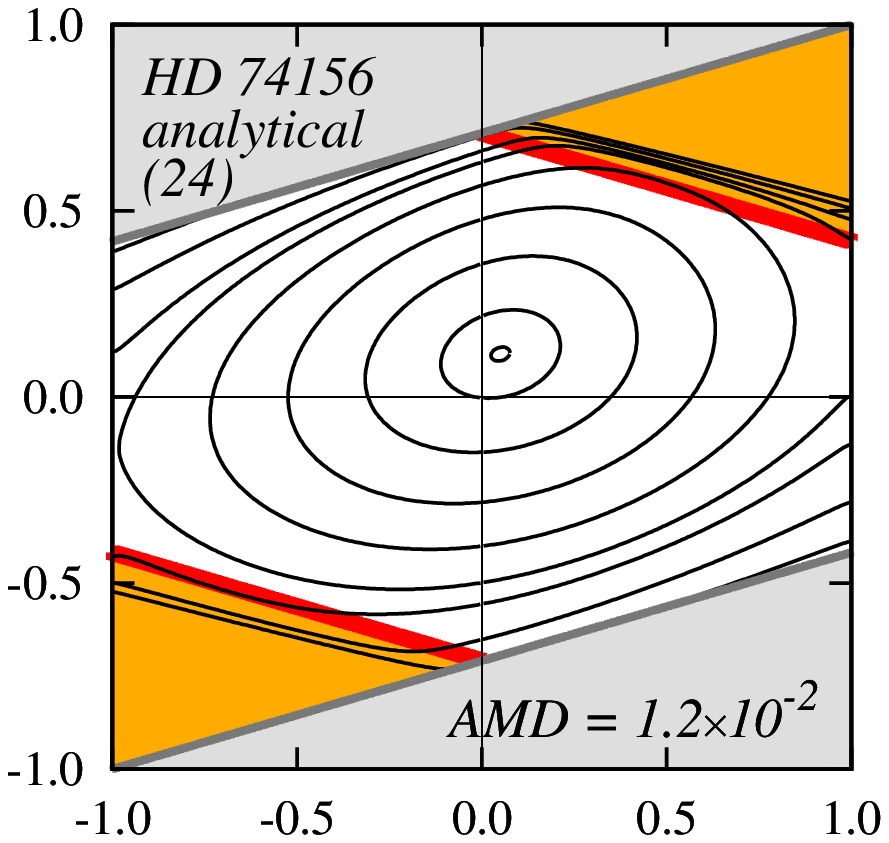}\hskip-3mm
          \includegraphics [width=45mm]{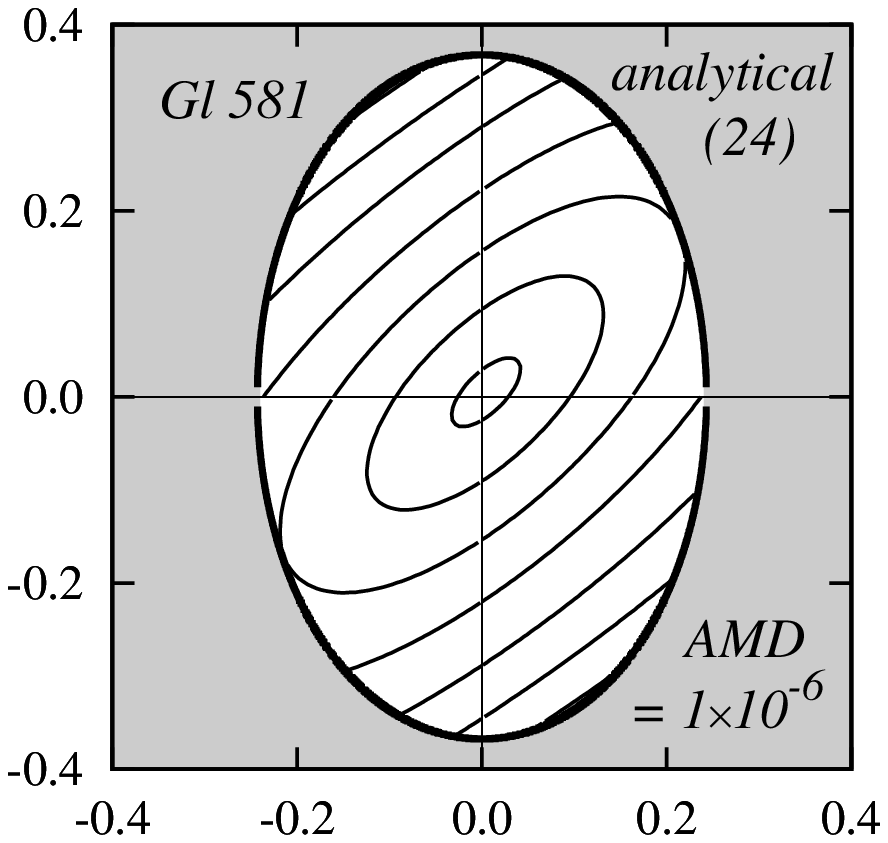}\hskip-3mm
          \includegraphics [width=45mm]{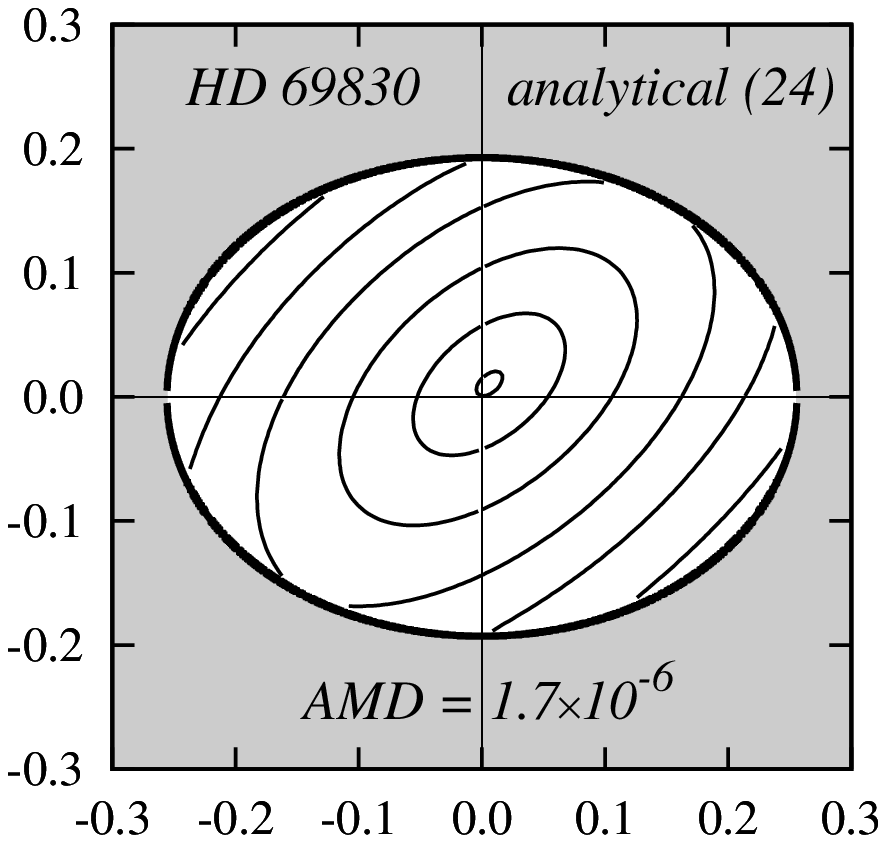}}
    \hbox{\includegraphics [width=45mm]{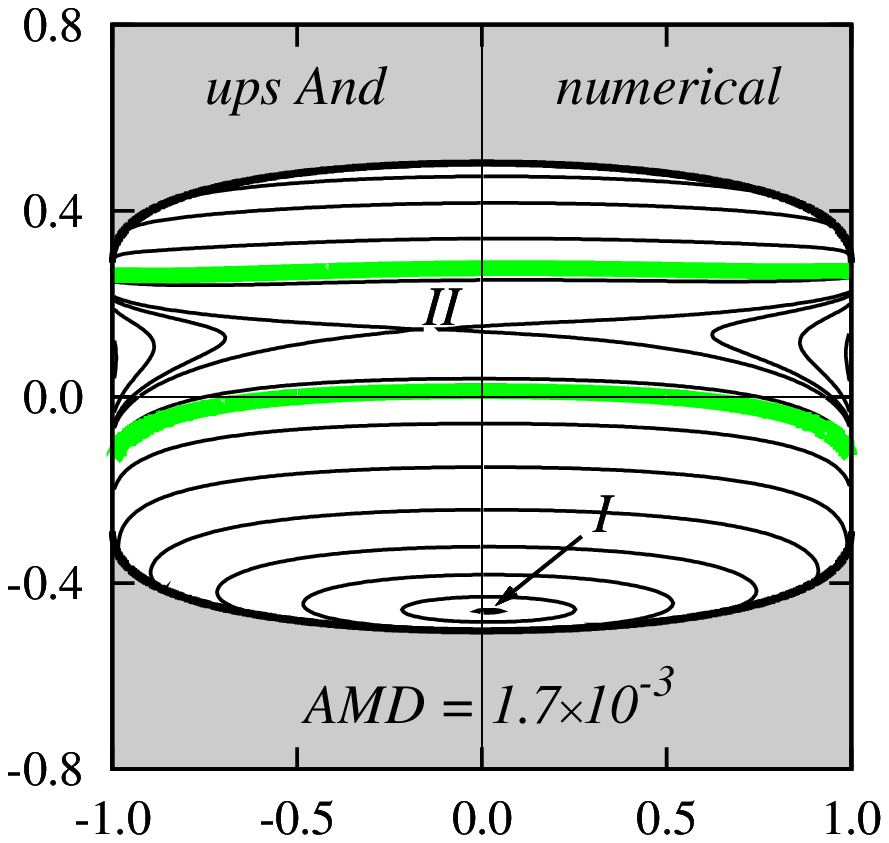}\hskip-3mm
          \includegraphics [width=45mm]{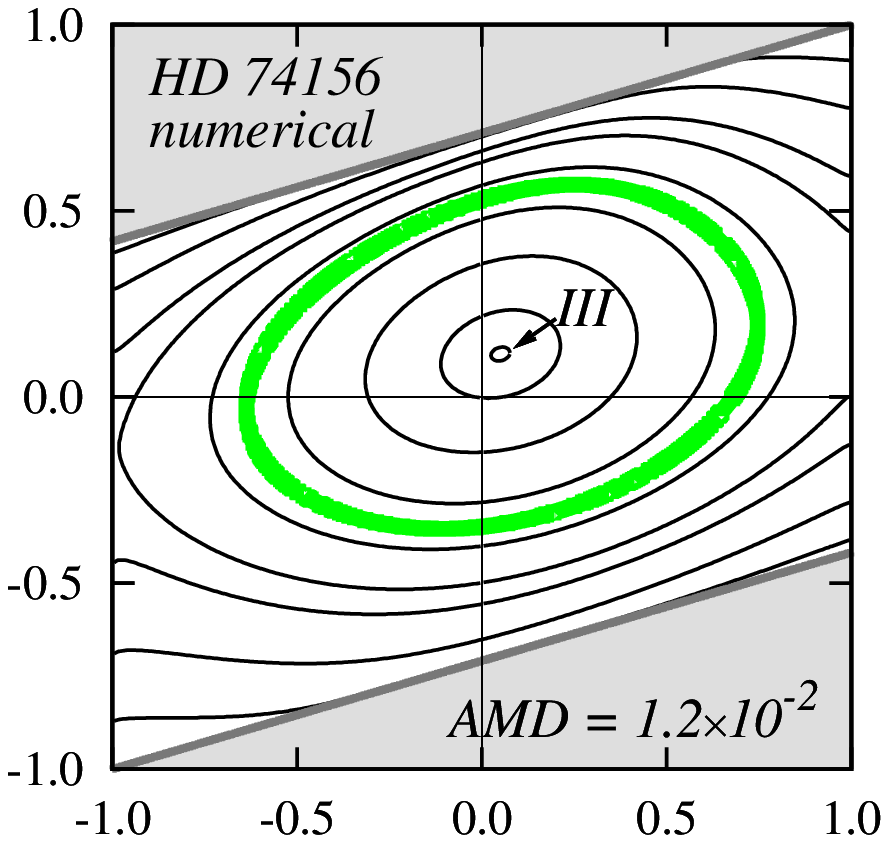}\hskip-3mm
          \includegraphics [width=45mm]{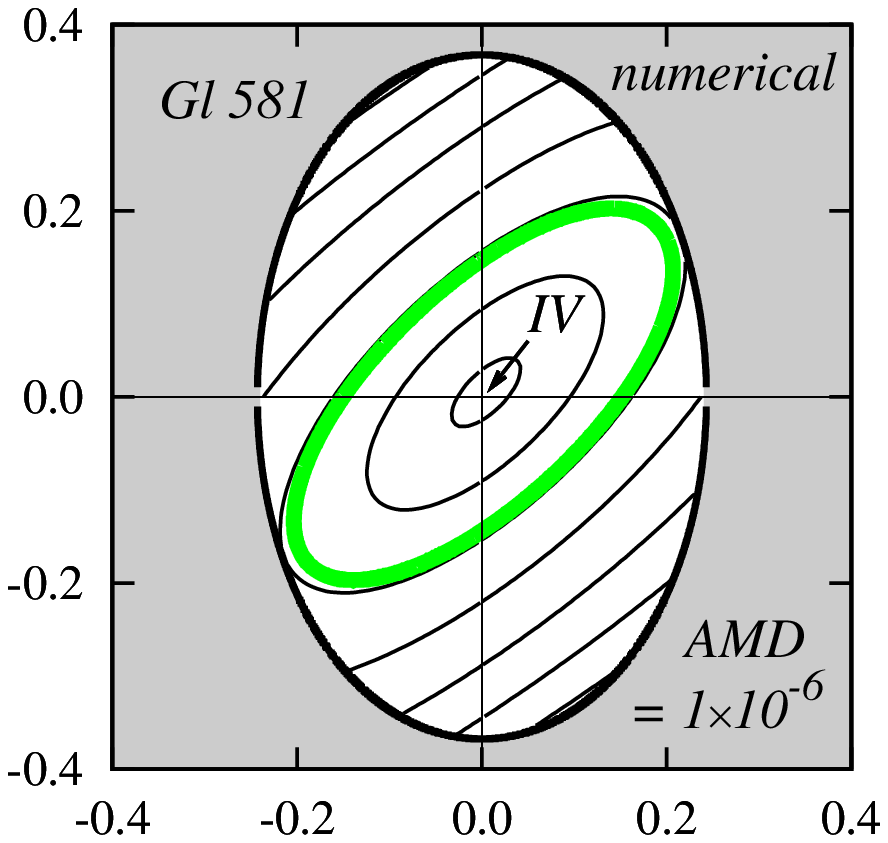}\hskip-3mm
          \includegraphics [width=45mm]{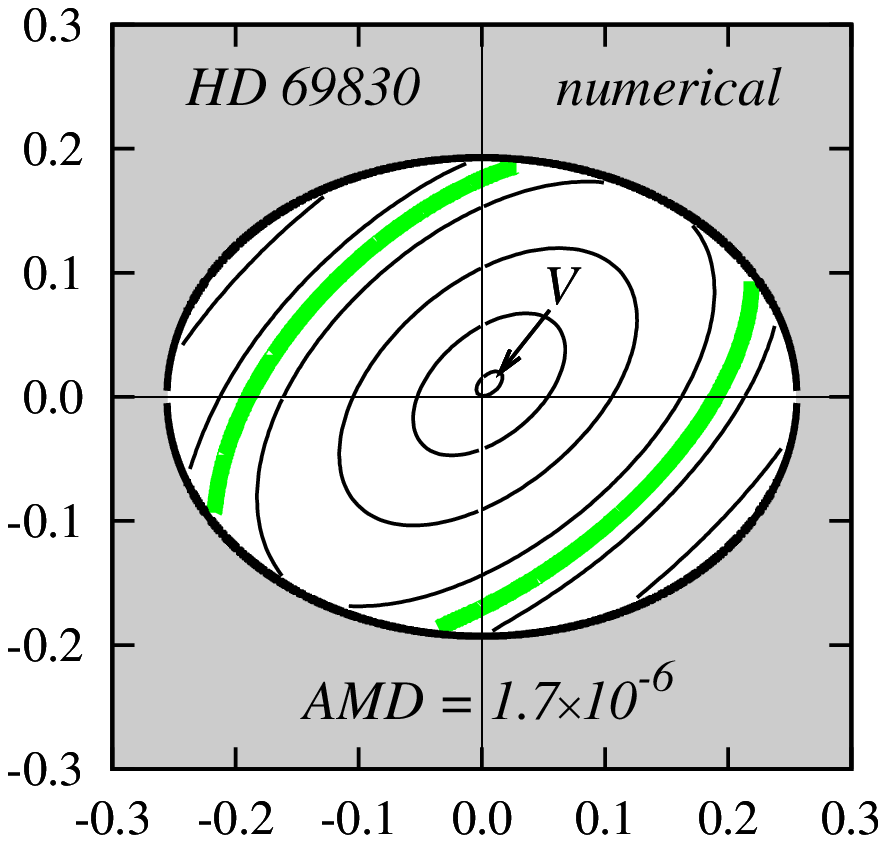}}
}}	  
 \caption{
 The secular energy levels on the \textit{\em symmetric  representative plane}
for selected three-planet systems. Map coordinates are  $(x\equiv e_1
\cos{\sigma_1}$, $y \equiv e_2 \cos{\sigma_2})$,  where the secular angles
$\sigma_1,\sigma_2$ are $0$ or $\pi$.   Black thin lines are for energy levels
obtained with different methods: panels in the top row are  for the octupole
theory, panels in the middle row are for the 24th-order expansion derived in
this paper, and  panels in the bottom row are for the semi-analytical averaging.
In gray areas, $e_3<0$, hence the motions are not permitted The light-gray areas
are for the regions of collisions between the inner and the middle planet.  In
orange regions, the secular expansion diverges. The thick, red straight lines
mark the anti-collision lines between  the planets and indicate the true border
of the convergence of the secular expansion. Orbital parameters are taken from
Jean Schneider Encyclopedia, and are given in terms of tuples 
 $\mathbf{p}_i \equiv (m_0 [\mbox{M}_{\odot}],m_1 [\mbox{m}_{\idm{J}}],
 m_2 [\mbox{m}_{\idm{J}}],m_3 [\mbox{m}_{\idm{J}}],
 a_1 [\mbox{AU}],a_2 [\mbox{AU}], a_3 [\mbox{AU}], e_1, e_2, e_3)$ as follows:
  $\mathbf{p}_{\idm{ups And}} = 
 (1.27,0.69,1.98,3.95,0.059,0.83,2.51,0.029,0.254,0.242)$,
 $\mathbf{p}_{\idm{HD~74156}} =
 (1.24,1.88,0.396,8.03,0.294,1.01,3.85,0.64,0.25,0.43)$,
 $\mathbf{p}_{\idm{Gl 581}} = 
 (0.31,0.0492,0.0158,0.0243,0.041,0.073,0.25,0.02,0.16,0.2)$,
 $\mathbf{p}_{\idm{HD~69830}} = 
 (0.86,0.033,0.038,0.058,0.0785,0.186,0.63,0.1,0.13,0.07)$. 
 Each panel is labeled with $AMD$ (expressed 
 in standard units) calculated for the nominal configuration.
 The secular energy levels for the 
 nominal systems are marked with green curves.
 }
 \label{systems}
 \end{figure*}
The results are illustrated in Figure \ref{systems}. We selected four
three-planet configurations. Their orbital elements are taken from the
Extrasolar Planets Encyclopedia of Jean Schneider, following the most recent
determinations of the orbital solutions. Each column in Fig.~\ref{systems} is
for one particular system, i.e., for $\upsilon$~And \citep{Butler2006a},
HD~74156~\citep{Bean2008},  Gliese~581~\citep{Lovis2006a} and 
HD~69830~\citep{Udry2007b}, respectively. Different approximations to the
secular theory are illustrated in rows. The top row panels are for the energy
levels calculated with the octupole theory,  panels  in the middle row are 
derived from the expansion of $\Hsec$ of the $24$-order, and  panels in the
bottom row illustrate the energy levels computed with the numerical algorithm.
The later case may be regarded as the exact solution to the problem thanks to 
adaptive, high order Gauss-Legendre quadratures which we used to compute the double
integral over $\Hpert$. In each case, we fixed $AMD$ consistent with the nominal
parameters of the examined systems.

\subsubsection{$\upsilon$~Andromedae}
First, we are looking at the exact (numerical) phase plots. In the case of
$\upsilon$~And, we found two types of equilibria. The first one, marked with I,
is the global minimum of the secular Hamiltonian. Hence, according to the
Lyapunov theorem, this equilibrium is  stable (because the Hamiltonian can be
regarded as the positive definite  Lyapunov function).  The equilibrium marked
with~II is a saddle point, and is stable in the linear approximation. It can be
verified by solving the eigenproblem of the linearized equations of motion in
the neighborhood of the equilibrium. Now, we can compare the outcomes of the
analytic theories with the  results of exact, numerical algorithm.  Apparently,
the high-order analytical theory is fully compatible with the numerical theory.
The largest deviations between the secular energies are of the order of
$10^{-9}$. This accuracy is preserved  even for eccentricities $e_1$ close to
$1$. On contrary,  the octupole theory provides only a crude representation of
the phase space.  The phase plot constructed with the help of this theory also
reveals an equilibrium of type I, however, at place of equilibrium II, 
qualitatively different energy levels appear (see the top-left panel  in
Fig.~\ref{systems} with three ``false'' equilibria).

To locate the ``real'' system in the energy plot, we mark the level of the
secular energy computed for the nominal parameters of the system with green,
thick curve. It provides only a crude imagination where the system is located;
one should be aware that we are looking at the representative plane (hence
$\sigma_{1,2}$ are fixed at specific values), and we do not take into account 
the parameter errors. Still, the plot tell us that while variability of $e_2$ is
limited, $e_1$ may be varied in all permitted range of eccentricity.

\subsubsection{HD 74156}
In the phase space of the HD~74156 system,  we discover only one equilibrium
(labeled with III) in the regime of small eccentricities. It is related to the
global maximum of $\Hsec$, and it means that this solution is Lyapunov stable.
The $AMD$ of the nominal system permit eccentricities to reach large values,
hence they enter the regions in which the secular Hamiltonian expansion
diverges (see the explanation in Sect.~2.5). These regions are marked in orange color.  The region of permitted
motions is also bordered by two collision lines of orbits, defined implicitly
through $a_2(1 - e_2) = a_1 ( 1 \pm e_1 )$ and they are marked with gray, thick
lines. In this case the view of the phase plot varies with the order of
expansion (or the applied algorithm). The high-order analytic theory
reconstructs the phase plot for $e_2 \sim [0,0.6)$ (white zone) in almost  whole
permitted range, nevertheless, in the region of divergent expansion the phase
plot is wrong. In the case of octupole theory, we obtain only a crude
approximation of the structure of the phase space, and again the theory
introduces  artifacts (two saddle points and an extremum).

We also plot the energy levels of the nominal system (in the same manner as we
did for $\upsilon$~And). Its parameters would evolve along this level relatively
distant from the equilibrium close to the origin.

\subsubsection{Gliese 581 and HD 69830}
The phase space of systems Gliese~581 and HD~69830 are  quite similar. The
region of permitted motions is limited to relatively small eccentricities.  In
both cases, we have only one equilibrium close to the origin that is related to
the global maximum of the secular Hamiltonian, and therefore they are Lyapunov
stable.  For the Gliese~581 planetary system,  the accuracy of the $24$-order
secular expansion is not very good; at the borders of  permitted motion, this
accuracy is at a level of $10^{-3}$ only.  For the third-order theory, this
accuracy is even worse, $\sim 10^{-2}$. In the case of HD 69830, the relative
accuracy of the high-order expansion is not worse than $2.5 \times 10^{-9}$.

\subsection{Secular dynamics of HD~37124}
As a particular system to study, we choose  the three-planet system of
HD~37124. It has been discovered by \cite{Vogt2005}. Remarkably,  the most
recent best fit solutions  to the observations are consistent with configurations involving
sub-Jupiter companions in orbits with moderate eccentricities. The eccentricity
of the outermost companion is not well constrained, nevertheless extensive
dynamical analysis of the RV data in \citep{Gozdziewski2007} make it possible
to locate this planet in a region between 8:3 and 11:4~MMRs with the middle 
companion. In that case, the orbital parameters can be regarded as well fitting
the assumptions of the secular theory.

 \begin{figure*}
 \centerline{
 \vbox{
    \hbox{\includegraphics [width=54mm]{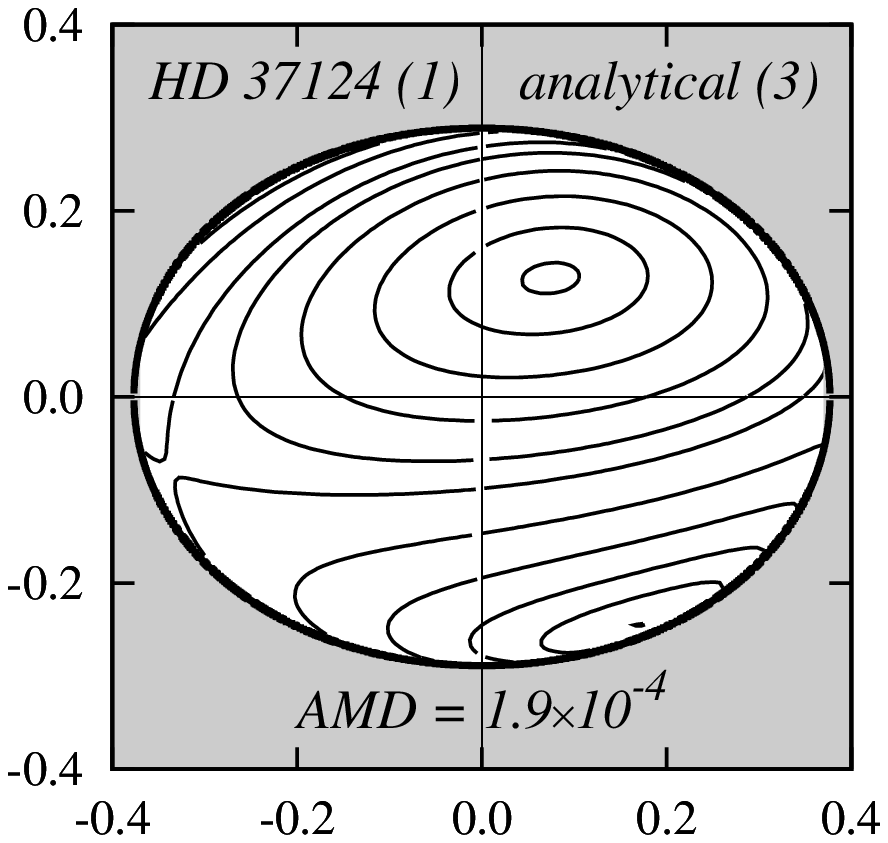}\hskip-2mm
          \includegraphics [width=54mm]{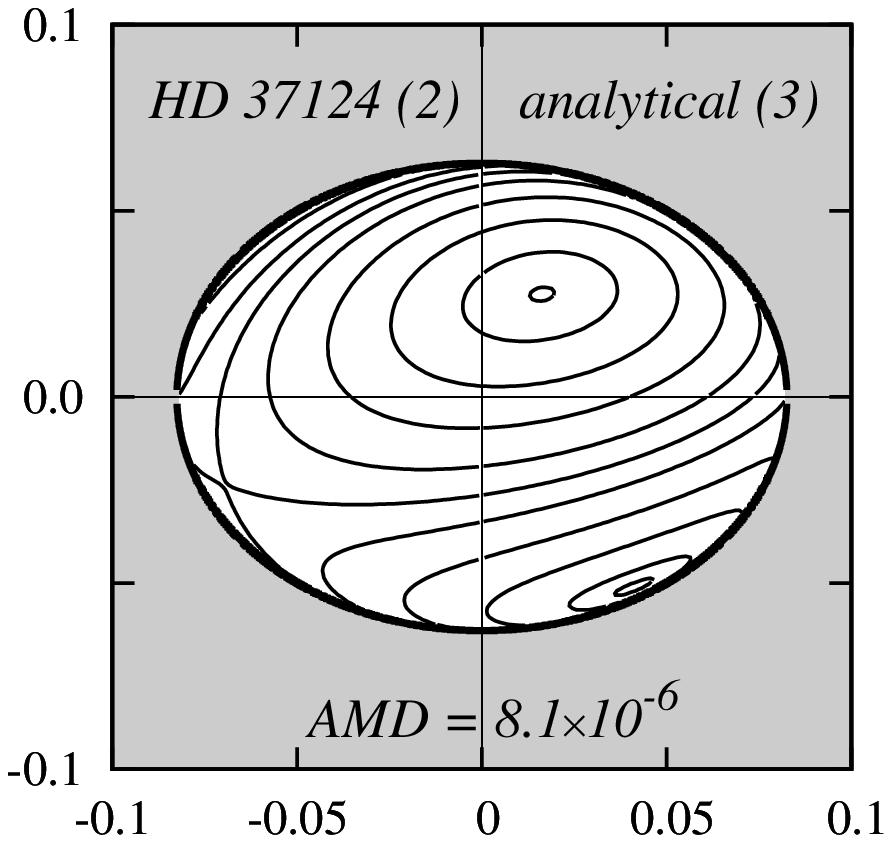}\hskip-2mm
          \includegraphics [width=54mm]{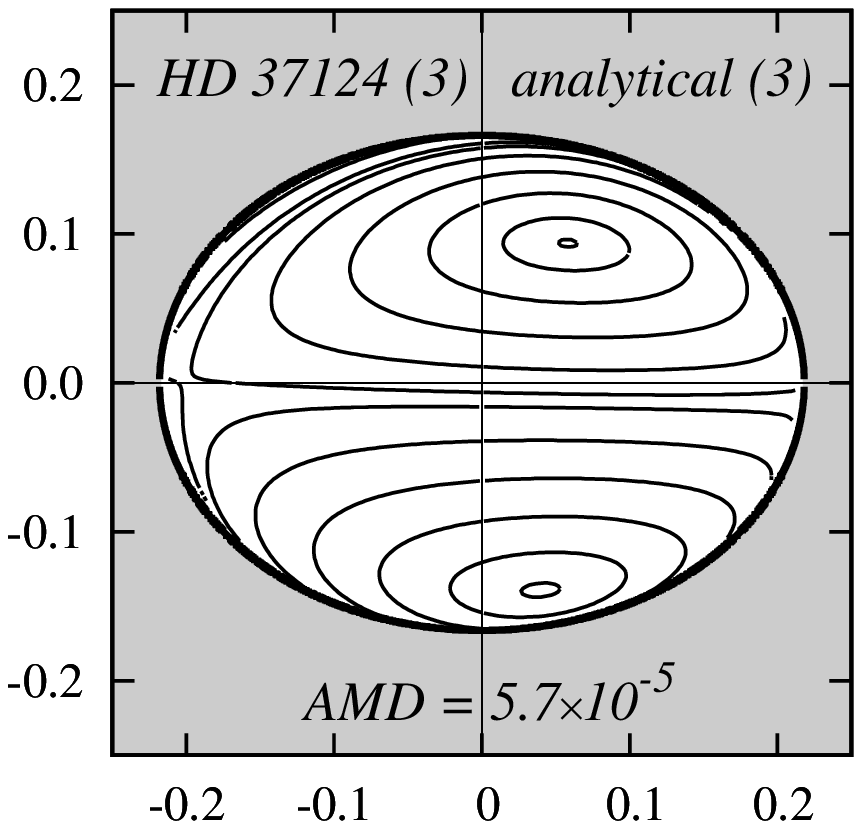}}
    \hbox{\includegraphics [width=54mm]{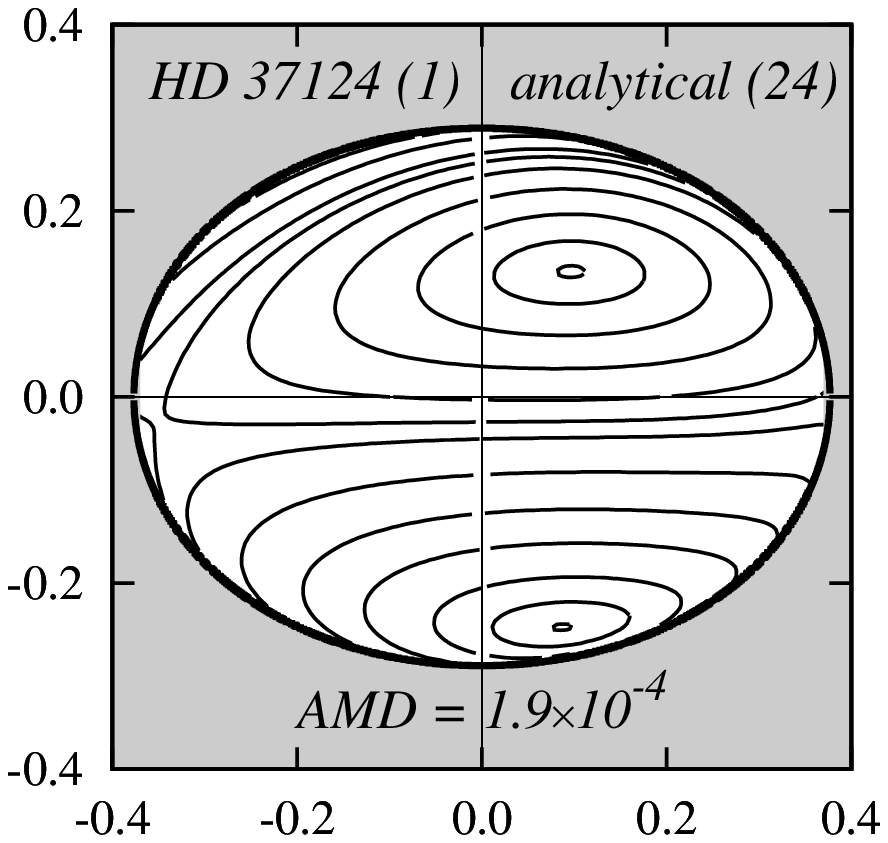}\hskip-2mm
          \includegraphics [width=54mm]{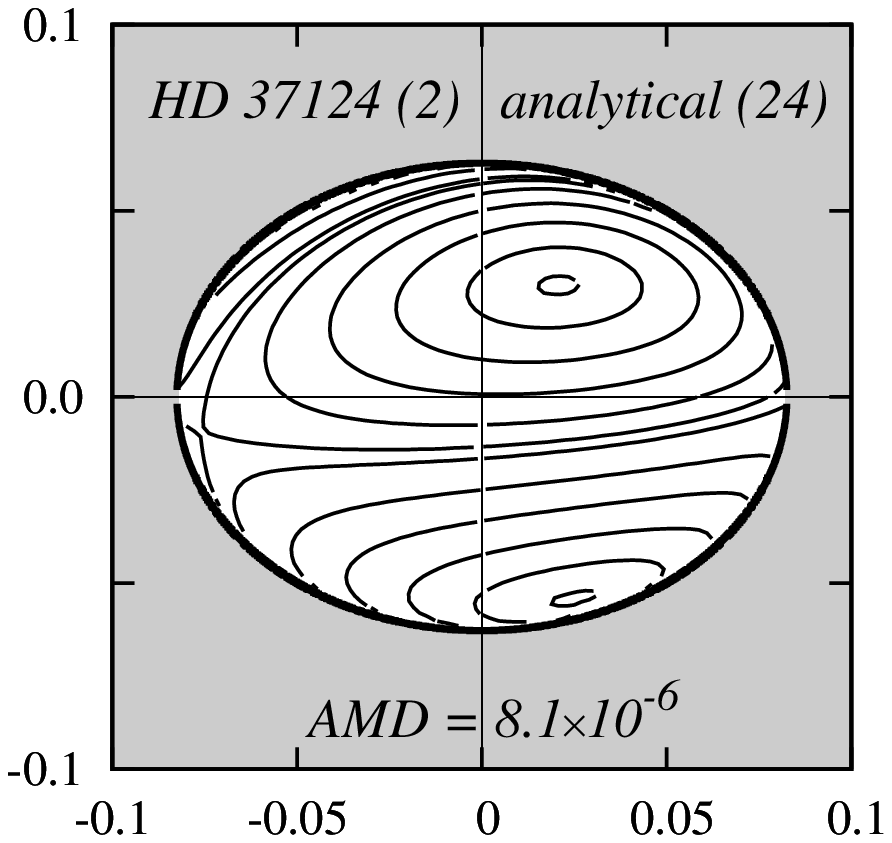}\hskip-2mm
          \includegraphics [width=54mm]{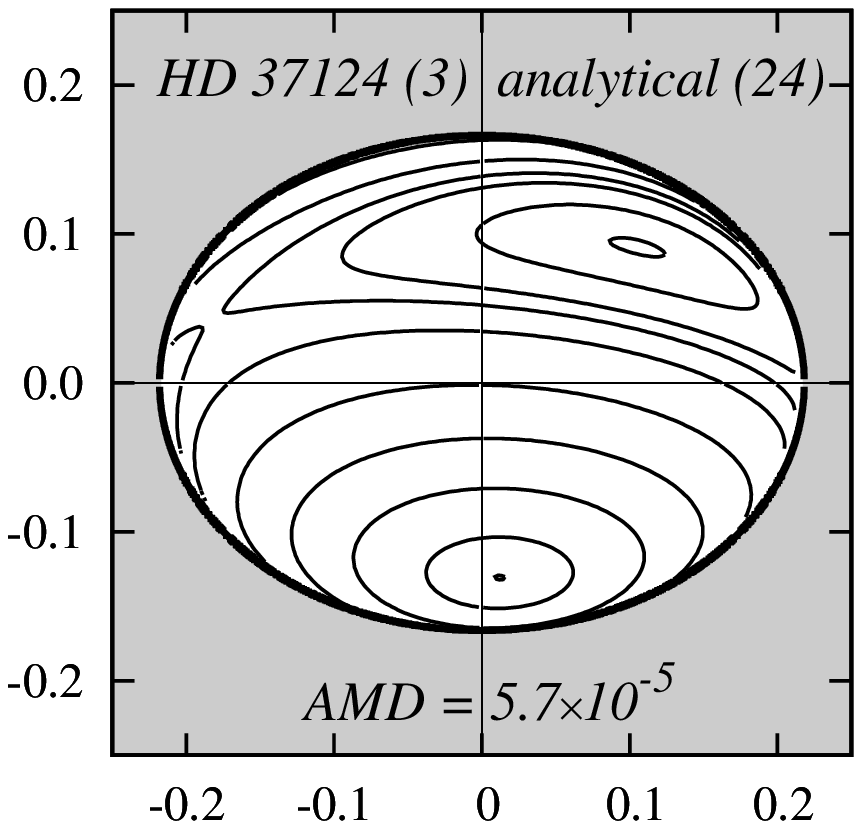}}
    \hbox{\includegraphics [width=54mm]{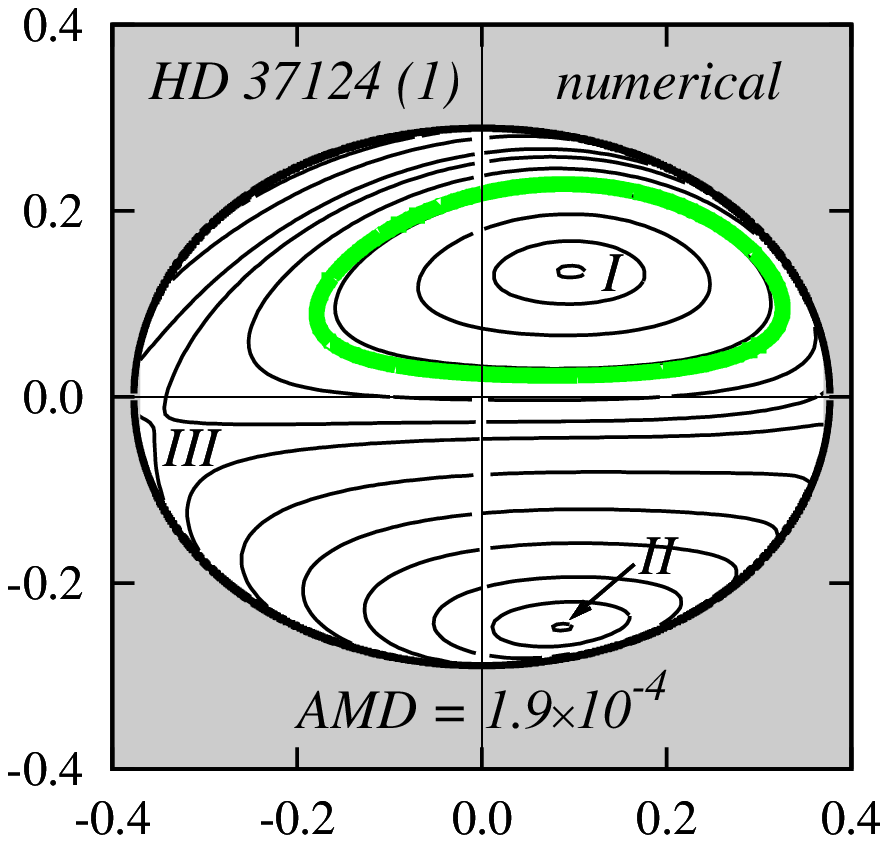}\hskip-2mm
          \includegraphics [width=54mm]{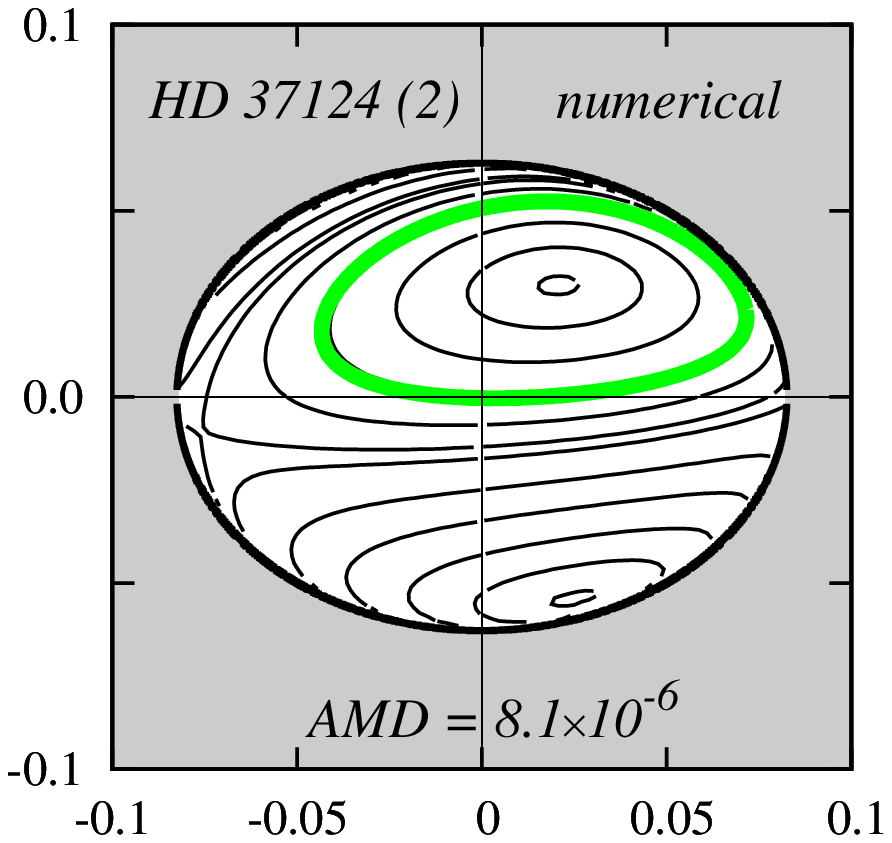}\hskip-2mm
          \includegraphics [width=54mm]{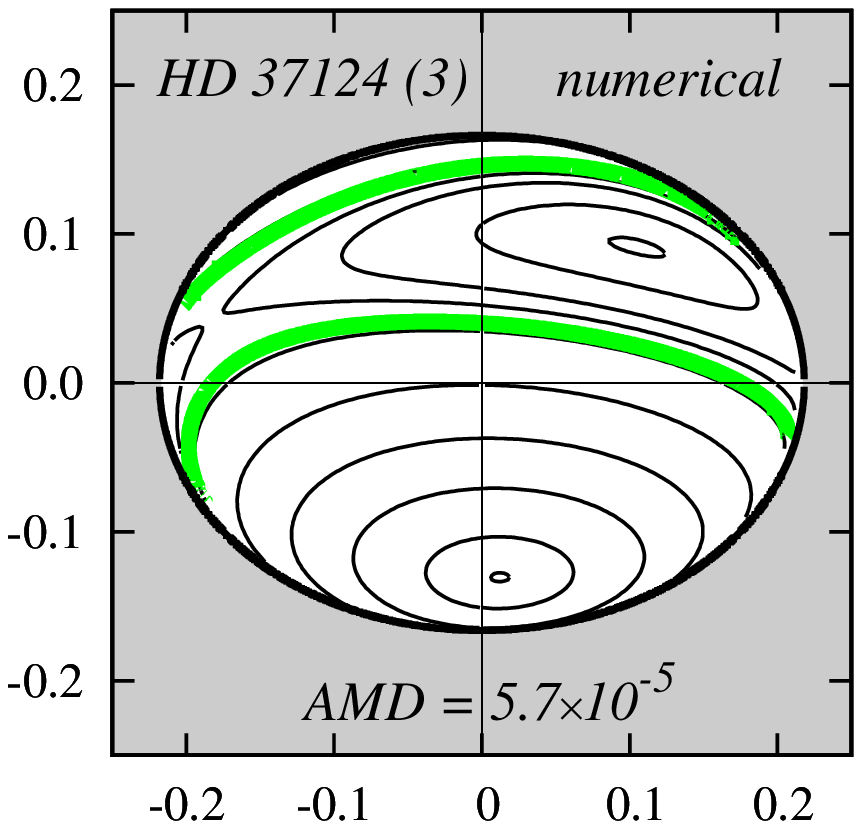}}
         }}
 \caption{
 The secular energy levels on the {\em symmetric  representative plane} for
three-planet system~HD37124 (see the text for more details). The  map
coordinates are $(x \equiv  e_1 \cos{\sigma_1})$ and $(y = e_2
\cos{\sigma_2})$,  where the secular angles are fixed at $0$ (the top
half-plane/the right half-plane) or $\pi$ (the bottom half-plane/the left
half-plane). The boundary between white and grey regions is for $e_3=0$ (i.e.,
the motion is permitted only in the white areas).  Black solid lines are for 
the secular energy  levels obtained with the help of the octupole theory (panels
in the top row),   by the expansion of the $24$-th order (panels in the middle
row)  and  with the semi-numerical averaging (the bottom row of panels). 
Orbital  parameters are taken from the discovery paper (Vogt et al., 2005)
(panels in the left column marked with 1), and from Go\'zdziewski et al. (2007)
(panels in the middle and in the right column, marked with~2 and~3,
respectively).  These parameters, in terms of tuples
 $\mathbf{p}_i   \equiv (m_0 [\mbox{M}_{\odot}],m_1 [\mbox{m}_{\idm{J}}],
 m_2 [\mbox{m}_{\idm{J}}], m_3 [\mbox{m}_{\idm{J}}],
 a_1 [\mbox{AU}], a_2 [\mbox{AU}],
 a_3 [\mbox{AU}], e_1, e_2, e_3)$ are the following: 
 $\mathbf{p}_1 = (0.91,0.61,0.6,0.683,0.53,1.64,3.19,0.055,0.14,0.2)$,
 $\mathbf{p}_2 = (0.78,0.624,0.606,0.581,0.519,1.632,3.212,0.037,0.003,0.048)$,
 $\mathbf{p}_3 = (0.78,0.650,0.584,0.567,0.519,1.668,2.740,0.091,0.040,0.132)$.
 Stationary solutions are labeled with I, II, and~III. Green curves
 are for the secular energy level of the respective nominal initial condition.
}
\label{hd37124}
\end{figure*}
 
Figure~\ref{hd37124} is for the energy levels in the  \textit{symmetric}
representative plane computed and calculated for three slightly different
orbital configurations related to possible orbital best-fits (panels in each 
column are for one orbital fit). Osculating elements of these configurations are
quoted in the caption to this figure. The first, kinematic solution to the
three-Keplerian model of the RV, is taken from the original discovery paper
\citep{Vogt2005}. Other two  best-fits are from \cite{Gozdziewski2007}.  The
energy levels are computed for fixed $AMD$  calculated for each particular
initial condition.

In the top row of Fig.~\ref{hd37124} we show energy levels calculated from the
octupole theory, plots in middle row are derived from the expansion of $\Hsec$
of the $24$-order, and the bottom row illustrates the results derived from the
numerical algorithm. In all cases, the analytical high-order theory is in
excellent agreement with the  numerical,  exact theory.  We checked that the
magnitude of largest, relative deviations between the analytic and numerical
results are of the order of $2.5 \times 10^{-6}$. On the other hand, the
octupole theory gives  relatively precise insight into the structure of the
phase space. All qualitative  features  of the energy plane are reproduced 
quite well.

The HD~37124 seems to be the most interesting example of the secular dynamics in
the real system found in this paper.  The energy planes reveal unusual dynamical
structures related to the equilibria in the secular system.  We know already
that they can appear as extrema as well as saddle points in the representative
plane.  For the same $AMD$, we can have three types of stationary solutions. Two
of them are characterized by extrema of $\Hsec$ in the four-dimensional phase
space: the first one is the maximum which  appears in the quarter with
$\sigma_{1,2}=0$, and there is  the global minimum of $\Hsec$ in the quarter
with $\sigma_{1}=0$ and $\sigma_{2}=\pi$. We also found  saddle points of
$\Hsec$ in the quarter with  $\sigma_{1}=\pi$, $\sigma_2=0,\pi$. Stationary
points  marked with I and II in Fig.~\ref{hd37124} are related to extrema of
$\Hsec$, hence they are stable. By examining the eigenvalues of the linearized
equations of motions in the neighborhood of the saddle point (equilibrium~III),
we checked out that  it is linearly stable. It can be localized in the
half-plane of  $\sigma_2=0$ or $\sigma_2=\pi$, depending on selected orbital
parameters. 

All these solutions appear in the range of moderate eccentricities, and in fact
can be located close to the actual positions of the best fit solutions. The
structure of the energy plane is also robust with respect to small changes of
the orbital parameters. In each panel, similarly to the previous systems, we 
mark the energy level of the respective nominal configuration  with the  green
thick curve. Curiously, depending on the chosen fit, the nominal system can
evolve in the quarter of the representative plane characterized by librations of
$\sigma_{1,2}$ around $0$ (the top-right quarter), or librations of $\sigma_{1}$
around $0$ and $\sigma_{2}$ around $\pi$ (the bottom-right quarter), as well as
$\sigma_{1}$ around $\pi$ while $\sigma_{2}$  can be librating around $0$  or
$\pi$ (the top-left or the bottom-left quarter). It means, that the apses of 
two innermost companions can be all aligned with the apsidal line of the
outermost planet,  they can be also anti-aligned or the apsidal directions can
be mixed.

The presence of these stationary solutions can be interpreted  in terms of the
two-planet theory. We recall that for the case of two planets, mode~I (with
apsides aligned)  corresponds to the maximum of the secular energy, while
mode~II (apsides anti-aligned) corresponds to the minimum of $\Hsec$. In the
case of three planets, we add the secular energies of three pairs of interacting
planets. Hence, in a region of the representative plane where  the maxima of
$\Hsec$ of these three planetary pairs can roughly coincide, we can obtain the
maximum of the total energy; by adding $\Hsec$ in the region where the
particular minima are close enough in the parameter space, we can obtain the
global minimum of the energy, and in the case of superimposed minimum and other
maximum we can obtain the saddle of the total energy. Geometrically, the
equilibria can be interpreted as combinations of  the secular modes known from
the theory of nonresonant two-planet system. For instance, the maximum of
$\Hsec$ can be related to triple mode~I (i.e., the neighboring solutions are
characterized with librations $\Delta\varpi$ around~$0$ for all  pairs of
planets), and the saddle point of $\Hsec$ is obtained for a superposition of
mode~I for some pair(s) and of mode~II for the other pair(s).  It is not clear
for us yet, what would mean a combination with  the NSR mode, in the regime of
large eccentricities (i.e., in the region of nonlinear-secular resonance).
Likely, it could be related to sophisticated secular dynamics.

\section{Conclusions}
The number of multi-planet extrasolar systems constantly grows. The Doppler
spectroscopy remains the most effective  detection technique. Unfortunately, the
measurements of RV are in some sense degenerate because due to symmetry of the
Doppler signal, we usually cannot determine the true inclination of planetary
orbits. Other parameters are usually determined with large uncertainties. Hence,
to characterize the dynamics of such systems we cannot rely only on single
initial conditions and effective, qualitative methods  of dynamical analysis are
very desirable. 

In this paper we consider the secular theory of a coplanar $N$-planet system
which is far from MMRs and orbital collision zones. In this
case, the high-frequency interactions can be averaged out and we obtain greatly
simplified picture of the long-term behavior of the system. This idea leads to 
the classic Laplace-Lagrange theory and its modern generalizations like the
octupole theory \citep{Ford2000,Lee2003}, high-order expansion of the secular
perturbation \citep{Henrard2005, Gallardo2005} or the semi-numerical  averaging
invented by \cite{Michtchenko2004}. 

Our work can be considered as a generalization of the octupole theory for
hierarchical triple systems characterized with large ratio of semi-major axes.
We have shown that in this case the perturbation can be averaged out over mean
longitudes with very basic change of integration variables that makes it
possible to express the integrand function as a polynomial of trigonometric
functions, without any need of relatively complex Fourier expansions.  To the
best of our knowledge, such method has been not applied in the literature.
However, during the final preparation of the manuscript we found a book of
\cite{Valtonen2006}, who use a similar idea to construct the octupole theory of
hierarchical triple stellar systems. 

Basically, the secular  Hamiltonian is expressed through the power series with
respect to the ratios of semi-major axes, without an explicit limitation on the
eccentricities. These series can be continued to practically any order. However,
if we apply the averaging algorithm presented in this work, the convergence
region of these series is usually limited. To avoid this problem, in this paper
we also propose  a further improvement of this method and to
generalize the expansion to the spatial problem. Our theory significantly improves the
octupole theory of two-planet systems. We have shown that it can be generalized
to any $N$-planet system fulfilling the assumption of the averaging theorem. The
simple ``trick'' of choosing the integration variables can be applied not only
for purely gravitational point-to-point interactions but also in  other models
in which  the mutual interactions can be expressed in powers of mutual distance
between objects in the system. For instance, now we work on  applying the
averaging algorithm to  relativistic and quadrupole moment perturbations
\citep{Migaszewski2008c}. Its generalization to the 3D problem (in particular,
for two-planet system) is also straightforward. Then it can be applied to the
study of secular dynamics in hierarchical triple-star systems or star--planet
configurations fulfilling  assumptions of the secular theory.

In this work, the secular theory is used to investigate stationary solutions in
the three-planet systems that are  relatively frequent in the known sample of
extrasolar planets. We found that the libration modes known in two-planet
configurations can be generalized for the multi-planet model.  Still, our study
of particular systems is quite preliminary and new, yet unknown stationary
solutions are expected to exist in this problem.

\section*{Acknowledgments}
We thank Tatiana Michtchenko for a detailed review 
and corrections that improved the manuscript.
This work is supported by the Polish Ministry of Sciences and Education, Grant
No. 1P03D-021-29. C.M. is also supported by Nicolaus Copernicus University
Grant No.~408A.
\bibliographystyle{mn2e}
\bibliography{ms}
\label{lastpage}
\end{document}